%% file: GGLLVM.tex
\documentclass[a4paper, 11pt]{article} 
\usepackage{amssymb}
\usepackage{color, graphics, graphicx}
\usepackage{amsmath}
\usepackage{amsthm}
\usepackage{mathrsfs}
\usepackage{fontenc}
\usepackage{lscape}
\usepackage{natbib}
\usepackage{authblk}
\usepackage{url}
\usepackage{bbm}
\usepackage{dsfont}
\usepackage{lscape}
\usepackage{setspace}
\usepackage{epstopdf}
\usepackage[shortlabels]{enumitem}
\usepackage{bigints}
\usepackage{rotating}
\usepackage{booktabs}
\usepackage{multirow}
 \usepackage[ruled,vlined]{algorithm2e}
\usepackage{algorithmic}

\RequirePackage[OT1]{fontenc}
\numberwithin{equation}{section}
\RequirePackage[colorlinks]{hyperref}
\hypersetup{
    colorlinks,%
    citecolor=blue,%
    filecolor=red,%
    linkcolor=blue,%
    urlcolor=blue
} 

\newcommand{\beq}{\begin{equation}}
\newcommand{\eeq}{\end{equation}}
\newcommand{\bea}{\begin{eqnarray}}
\newcommand{\eea}{\end{eqnarray}}
\newcommand{\ba}{\begin{array}}
\newcommand{\ea}{\end{array}}
\newcommand{\bi}{\begin{itemize}}
\newcommand{\ei}{\end{itemize}}
\newcommand{\ben}{\begin{enumerate}}
\newcommand{\een}{\end{enumerate}}
\newcommand{\nn}{\nonumber}

\renewcommand{\r}{\right}
\renewcommand{\l}{\left}
\long\def\symbolfootnote[#1]#2{\begingroup\def\thefootnote{\fnsymbol{footnote}}\footnote[#1]{#2}\endgroup}

\newcommand{\tr} {\text{tr}}

\newcommand{\bfth}{\boldsymbol{\theta}}
\newcommand{\bfS}{\boldsymbol{\mathcal{S}}(\boldsymbol{\theta})}

\title{ \Large \bf  
GLAMLE: inference for multiview network data  in the presence of latent variables,  with application to commodities trading}

\author[1,2]{C. JIANG}
\author[2]{D. LA VECCHIA}
\author[3]{R. RASTELLI}
\affil[1]{\footnotesize Department of Biostatistics,  Epidemiology and Informatics,  University of Pennsylvania;}
\affil[2]{\footnotesize Research Center for Statistics, University of Geneva;}
\affil[3]{\footnotesize School of Mathematics and Statistics, University College Dublin.}
\affil[ ]{chaonan.jiang@pennmedicine.upenn.edu; davide.lavecchia@unige.ch; riccardo.rastelli@ucd.ie}

\date{\today}

\begin{document}

\maketitle

\begin{abstract}\begin{footnotesize}
The statistical  analysis of import/export data is  helpful to understand the mechanism that determines exchanges in an economic network. The probability of having a commercial relationship between two countries often depends on some unobservable (or not easy-to-measure) factors, like  socio-economical conditions, political views,  level of the infrastructures.  To conduct inference on this type of data, we introduce a novel class of latent variable models for  multiview networks, where a  multivariate latent Gaussian variable 
determines the probabilistic behavior of the  edges. 
We label our model the Graph Generalized Linear Latent Variable Model (GGLLVM) and 
we base our 
inference on the maximization of the Laplace-approximated likelihood. We call the resulting M-estimator the Graph Laplace-Approximated Maximum Likelihood Estimator (GLAMLE) and we study  its statistical properties. 
Numerical experiments on simulated networks illustrate that the GLAMLE yields fast and accurate inference. 
A real data application to commodities trading in Central Europe countries unveils the import/export propensity that each node of the network has toward other nodes, along with additional information specific to each traded commodity. \end{footnotesize}
\end{abstract}

\section{Introduction}

\subsection{Motivating example} \label{realexamples}

In our daily life, we are involved in different types of complex systems, related to economics, social sciences, transportation, energy to name some examples. The data collected on those systems are of different nature: e.g. they can be  binary, count, continuous data or mixture thereof. These complex systems can be thought of as  networks; we refer to \citet{K09, LKR13, KC14} for book-length discussions on the statistical analysis of network data.

 Among the different types of networks, multiview networks \citep{Gollini2016246, Salter-Townshend20171217, angelo2019900} have been recently attracting the attention of the research community.  This type of networks  consists of multiple layers of interactions among the same nodes. For instance, \citet{DDNA15} study  multilayer networks, assuming that each layer represents one possible state of the system, namely a network state. 
 A multilayer network is obtained, with the nodes replicated along the different layers. This type of models are helpful in various fields, including biology (e.g. to conduct inference on the protein-protein interactome), and social sciences (e.g. to analyze social systems, where individuals can have political or economical relationships, or can interact using different communication channels, like face-to-face interactions,  phone calls, e-mail, or social media, like Twitter and Facebook). The methodological investigation of our paper is related to the analysis of a socio-economical multiview network, that  we introduce in this section to motivate our work. In Section \ref{Sec: application}, we  
re-consider the dataset in more detail. \\

\begin{figure}[hbtp]
\begin{center}
\includegraphics[width=1\textwidth, height=0.26\textheight]{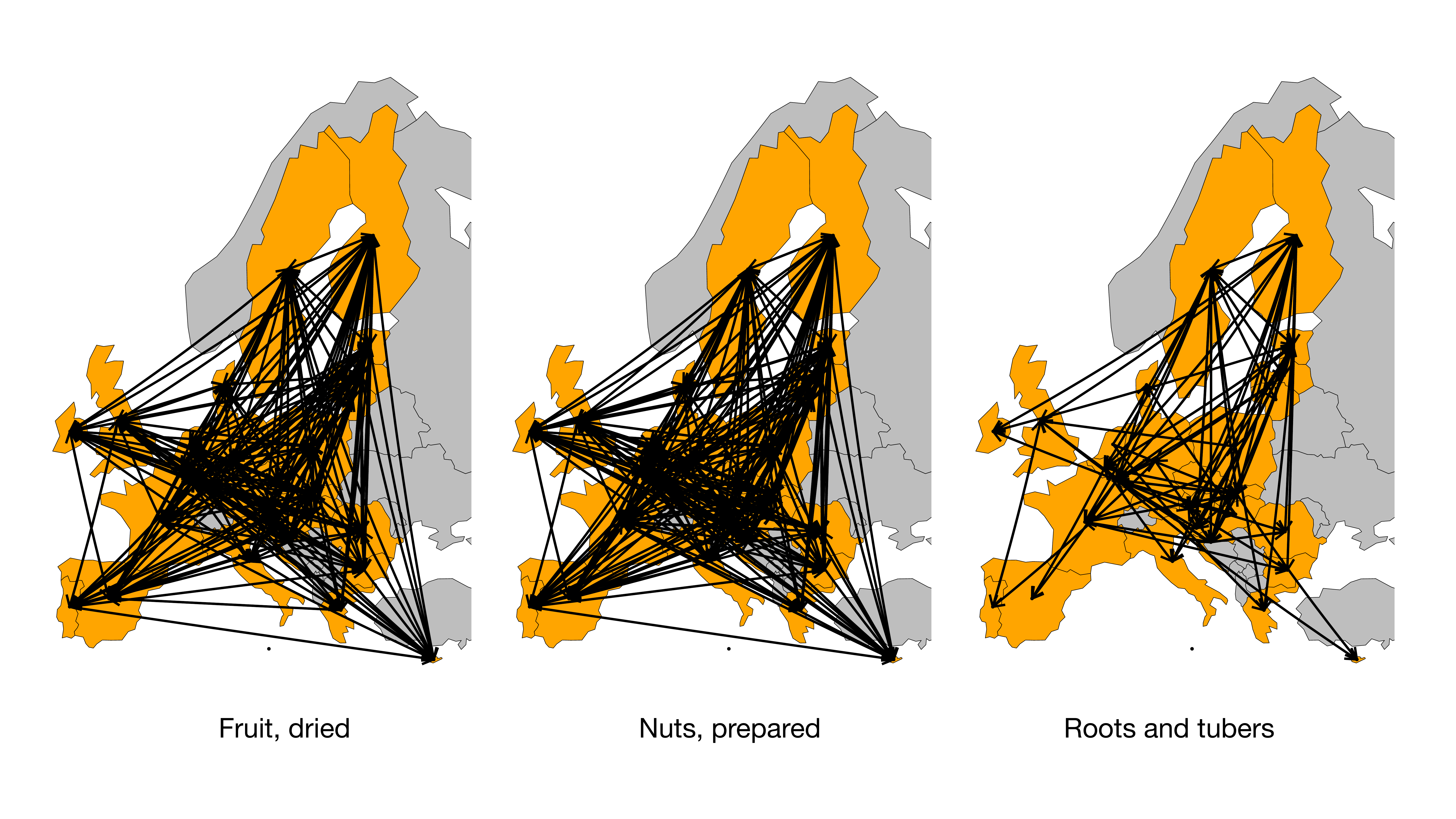}
\caption{Directed graphs for 3 different layers (associate to three commodities: fruit dried, nuts prepared and roots \& tubers) in the FAO data set in 2010.}
    \label{Fig: FAO}
\end{center}
\end{figure}

\noindent\textit{Data.}   The dataset introduced by  \citet{DDNA15} is an example of a socio-economical complex system. The dataset that we consider contains records on different types of trade relationships among 28 European countries, obtained from the Food and Agriculture Organization of the United Nations (FAO). The import/export network is an economic network in which layers represent commodities (primarily food products), nodes are countries, and the edges at each layer represent import/export relationships of a specific commodity among countries. The edges of each of the networks are directed. We investigate 
the presence of the trade relationships across the 28 European countries by simply defining the edge weight being equal to one when there exists a commercial exchange among two nodes, or zero otherwise (i.e. the data is binary). A multiview (also called multiplex) network is obtained by simultaneously considering the 364 layers, each corresponding to a possibly different binary network, and each corresponding to a different traded product.  This network dataset has been attracting the attention of both the empirical and theoretical research communities, becoming a benchmark for the analysis of multiplex network data; see, among the others, \citet{Ra_etal2018} and \citet{YQ21}.
Similar application contexts have been considered by other works in the networks literature, including \citet{sewell2016latent, d2020modeling, melnykov2021finite}.
Figure \ref{Fig: FAO} offers a graphical visualization of the directed graph associated to three selected layers: these plots illustrate that some commodities have more directed edges than others, and the distribution of edges can be quite diverse across layers. 
Moreover, the directed edges depict the fact that each country may be at the same time a \textit{receiver} for some commodities (e.g. an importer), and a \textit{sender} for the same or some other commodities (exporter). Therefore, these different facets in node's behaviors invite a sophisticated model which can break down and capture the driving factors for the data. 
The multiview network that we consider in this paper is a subset of the dataset analyzed by \citet{DDNA15} and available at \url{https://manliodedomenico.com/data.php}. \\ 

\noindent\textit{Statistical challenges.} 
In the statistical analysis of this dataset, it is relevant (e.g. for economic decision making) to have general information on the underlying probability mechanism that explains the presence of edges, bypassing the idiosyncrasies related to each traded item. The resulting inference is informative on the propensity that country $i$ has to set up economic relationship, import/export, with country $j$. At the same time, it is important to gain an understanding of the specific features that each layer has for import/export over the network. 
To characterize these facets, the combination of a Generalized Linear Model (GLM) and a latent variable model for network data (see next section for the key references) can be used to define a suitable statistical modeling and an accurate inference approach. 
One can conjecture that the edge probabilities depend on some variables  (e.g. socio-economical, demographical, political, cultural, technological, environmental) that affect consumers' behavior; we refer to \citet{WTO14} for a review of the World Trade Organization report on the factors shaping the world trades. Those factors cannot be easily measured by observable proxies and, in fact, they are latent variables. On the one hand, ignoring these variables can severely bias the estimates and negatively affect the inference conclusions that are obtained through GLMs. On the other hand, the presence of unobservable covariates entails the need for a marginalization procedure, which should yield the log-likelihood to be maximized. Unfortunately, this approach is either computationally very intensive (already with moderate to small number of edges) or unfeasible (even with a moderate to small number of latent variables). 
As a result,  standard inference based on
maximum likelihood estimation is not a viable inference method for the analysis of our network data. To cope with this issue, the extant literature relies on Markov chain Monte Carlo (MCMC) procedures or variational methods; see \citet{HRH02} and \citet{Gollini2016246} among the others for the logistic case. 
Although these procedures are able to overcome the difficulties related to the standard likelihood approach, the extant MCMC procedures and variational methods suffer from some other inference issues. For instance, MCMC methods are computationally demanding, whereas variational methods are faster but they can be biased and/or inaccurate; see Section \ref{Sec: VA} for numerical illustration. 

\subsection{Related work}


The recent research on statistical network analysis has brought novel inference procedures, whose aim is to explain the key features of network data.  The main goal consists in modeling the dyad variables $\{Y_{ij}\}$ associated to the pairs of nodes  $\{i,j\}$. Statistical models should be able to describe (or predict) the dyadic dependence structure over the network. However, this dependence can be exceptionally complex; see \citet{newman2018networks}. 

One popular class of network models is the so-called Exponential Random Graph Model (ERGM). See, e.g., \citet{FS86},  \citet{Wasserman_1996}, \citet{S02}, \citet{RO07} and \citet{LKR13} for a book-length presentation. Those models are available for either undirected or directed graphs: some network statistics (like e.g. the number of edges, triangles, stars) and  
some observable variables explain the dependence among network data. 
These models try to capture the main features of networks within a unified framework. However, ERGMs are characterized by intractable likelihood functions, which lead to heavy computational cost, so they often become impractical. 

Working on the limitations of ERGMs, \citet{HRH02} notice that, in some (social) network data, the probability of a relational tie between two individuals may be higher when the characteristics of the individuals are similar. A set of individuals with a large number of mutual social ties may be indicative of the presence of a cluster in a suitable latent social space of characteristics. Starting from this consideration,  \citet{HRH02} propose a modeling framework which relies on a latent space representation, whereby the nodes of the network are characterized by their individual latent coordinates, which in turn determine their connectivity patterns. The resulting class of models is rather flexible: the two main resulting approaches are called the latent distance model (which considers the Euclidean distance between pairs of nodes to determine the presence of relational ties) and the latent projection model (which uses the dot product).  

In the network literature, latent variable models (LVMs) are flexible statistical tools, able to capture relevant data features of interest and that include the models of \citet{HRH02} as a special case. A  formal analysis of these models is provided by  \citet{RFR16}, who study the properties of a family of LVMs for networks, including both the distance model and the projection model. The authors show that, under suitable conditions, the latent variable framework can represent the heavy-tailed degree distributions, positive asymptotic clustering coefficients, and small-world behaviors that often occur in observed social networks.
In the last two decades, LVMs have been employed for a variety of applications in different fields
\citep{Durante20171547, Chiu2014139, westveld2011mixed, Tafakori2021, Jin2019236}, and have been extended and adapted for the analysis of various network structures, including dynamic networks \citep{Sarkar20051145}, multiview networks \citep{Gollini2016246}, bipartite networks \citep{Friel20166629} and ranked networks \citep{Gormley200790}.
For a review, see e.g.  
\citet{Salter-Townshend2012243, RFR16, Raftery20171531, Smith2019428}. 

In the statistical literature, LVMs are popular too. As explained in \citet{BKM11}, one common use of LVMs in statistics is to reduce dimensionality. Large-scale enquiries are a typical example:  surveys contain so many variables that it is difficult to see any pattern in their interrelationships. Nevertheless,  many of those variables convey similar info---for instance, in socio-economical surveys the variables may reflect political and/or economical views. LVMs provide a  way to condensate similarly informative variables. Another statistical use of LVMs  is in those situations where researchers need to model concepts that are not directly observable (they are latent) and the statistical procedures require to combine manifest and latent variables. Examples of the mentioned inference issues are available in psychology (educational testing to study mental abilities, see e.g. \citealp{Jo69}) and in economics (to study the welfare, see \citealp{Sen87} for an economic overview and \citealp{HRVF04} for a statistical perspective). Over recent years, the 
Generalized Linear Latent Variable models (GLLVMs, see \citealp{SRH04,BKM11}) have been gaining considerable attention for these problems. The key GLLVM feature is that they start with the basic and easy interpretable GLM structure as in \citet{McN89}, but they enhance it by assuming that the canonical parameter can be expressed as a linear combination of latent variables (thus latent factors are considered, as in LVMs).  Thanks to this property, GLLVMs are useful statistical tools for the analysis of multivariate and high-dimensional data in many disciplines, like e.g. economics \citep{HRVF04}, finance (see \citealp{HuSVF09}) or ecology \citep{O16}.  Inference on the model parameters is typically conducted via MCMC, 
variational or Laplace approximation; see  \citet{HRVF04},  \citet{HWOH17} and  \citet{Nikuetal2019}.

\subsection{Our contribution}

This paper aims at combining two different (seemingly unrelated) strands of literature: the one about LVMs for networks
 and the one about GLLVMs for multivariate random variables. We build on the theoretical and methodological aspects pivoting on the GLM structures which are in common across these two areas of research. The result is a new methodology which combines a new useful modeling framework and a computationally efficient inference procedure. We provide below more details on our original contributions, which are four-fold.

(i) \textit{Modeling}. We introduce of a novel class of LVMs for the statistical analysis of network data. 
We focus on a variant of the so-called projection model of \citet{HRH02}, which relies on the dot product between latent variables. Specifically, our modeling combines, for the first time in the statistical and network literature,
LVMs \citep{HRH02} and the GLLVMs \citep{BKM11}. The resulting model formulation is original, however it can be connected with other works already available in the literature; in particular, we note interesting modeling similarities with the works of \citet{Gollini2016246}, \citet{Durante20171547} and \citet{d2020modeling}. 
We describe and study our model under general assumptions, so that we are able to take into consideration both binary and non-binary networks.
On the one hand, our latent projection model approach benefits from a mathematically very tractable definition (e.g. as in \citealp{hoff2005bilinear, hoff2021additive, Durante20171547}), which is essential for the derivation of our methodology and for its numerical implementation. On the other hand, the interpretation of our latent factors is different from the traditional Euclidean distance models and perhaps it is more challenging. Our methodology relies on an adaptation of the GLLVMs (as derived in \citealp{HRVF04}) to multiview networks, for the statistical analysis of multivariate data. 
We call our class of models the Graph Generalized Linear Latent Variable Models (GGLLVMs), to emphasize the role of graphs (needed to mathematically model the network) and of the GLMs in the presence of latent variables  (needed for statistical purposes).  See Section \ref{Sec: Assum} for a more formal definition of our model.

(ii) \textit{Fast and accurate inference}. The Laplace approximation \citep{HRVF04} is an effective method for the implementation of GLLVMs. In particular, it provides fast and accurate inference in the setting of multivariate data. In this paper, we show that similar advantages can be obtained also for GGLLVMs in the setting of multiview networks. To estimate the GGLLVMs parameters, we introduce a Laplace approximation to the log-likelihood computed over the network views. This yields a new estimator that we call the Graph Laplace-approximated maximum likelihood estimator (GLAMLE). 
To the best of our knowledge, none of the extant LVMs for network data  makes use of the Laplace approximation to conduct parametric inference. We study the key theoretical aspects of the GLAMLE, showing its statistical guarantees; in Sections \ref{Sec: GLAMLE}-\ref{Sec: Infe}. 

(iii) \textit{Implementation and computational aspects}. We take care of the numerical aspects related to the implementation of the GLAMLE. Our computational approach contrasts with most of the literature on statistical network analysis, which typically considers a Bayesian framework, and relies on MCMC methods to obtain posterior samples of the model parameters; see \citet{HRH02}, \citet{Rafetal12}, \citet{rastelli2018computationally}. Because of the simulation and estimation steps, the current state of the art Bayesian methods require very intensive computation. Variational inference \citep{SM13, Gollini2016246} offers a viable alternative, though it is sometime difficult to ensure consistent and/or accurate estimates. 
 In Section \ref{Sec: MC_gllvm}, we provide extensive simulation studies to 
illustrate the accuracy and the speed
of our estimation method. 
We give significant evidence that the GLAMLE yields less biased and more precise estimates than variational methods, but at a comparable computational cost. This makes our estimation method highly competitive: it can better address the computational gap arising in literature on network data.

(iv) \textit{Real data analysis}. In Section \ref{Sec: application}, we discuss  the FAO real data motivating example. We illustrate the model-based summaries that our novel approach can yield, providing novel insights on the economic network. Our analysis gives new visualizations and interpretations, breaking down some of the driving features which define this dataset. The proposed GGLLVM approach offers also the possibility to have easily interpretable diagnostic tools, which complete the data analysis.

\section{Network graphs} \label{Sec: Assum}

\subsection{Notation and assumptions}

We introduce our notation and define various latent variable models for networks. Models will differ on the assumptions on the probabilistic structure of the underlying stochastic process. The first step is to model the network via a graph. To this end, we introduce:

\textbf{A1.} Let $\mathcal{G} = (V,E)$ be a random graph, where $V$ is the set of node labels and $E$ is the set of random edges. Without loss of generality, we label the nodes with $V = \{1,...,n_{V}\}$, and so $\vert V \vert = n_{V}$ is the number of actors in the network. Since we do not consider self-edges among the nodes, the number $m$ of dyads in the network is given by
\beq  \label{Eq: m}
m=
\begin{cases}
  {n_{V}(n_{V}-1)}/2              \quad      & \text{for undirected relations }\\ 
  {n_{V}(n_{V}-1)}                \quad      & \text{for directed relations } 
  \end{cases}.
\eeq
We use  $i \sim j$, for undirected relations and $i \to j$, in the case of directed relations, $\forall i,j \in V$ with $i\neq j$.

We consider a random field taking values on the edges of a random graph $\mathcal{G}$ and we label it $\mathcal{{Y}}=\{ Y_{ij}, (ij)\in \{ V \times V\}\}$. The field $\mathcal{{Y}}$ is directly observable by the experimenter, through {realizations} of $\mathcal{G}$, and its values are influenced by a latent variable $ \boldsymbol{Z} \in \mathbb{R}^q$, for $q \ll n_V$. 
We label this type of models as LVM for networks. 

In the statistical literature, $\mathcal{G}$ is defined often  in terms of the nodes and the corresponding measurements of $\mathcal{{Y}}$ on pairs of nodes. In the simplest cases, $Y_{ij}$ is a dichotomous (Bernoulli) random variable indicating the presence ($Y_{ij}=1$) or absence ($Y_{ij}=0$) of some relation of interest  for $i \sim j$ or for $i \to j$. The data are represented by an $n_V \times n_V $ matrix $\boldsymbol{Y}$, called sociomatrix or adjacency matrix. For undirected relations, the adjacency matrix is symmetric, whilst for directed relations  it can be asymmetric. Nodes in the network may represent individuals, organizations, or some other kind of entity. Edges correspond to the social links or commercial exchanges between the units, and they may be directed 
or undirected.

As remarked by \citet{HRH02}, in some social network data, the probability of a relational tie between two individuals may depend on the same latent characteristics of the 
individuals, which make them similar in some sense. 
In principle, the latent variables which characterize/span this space can be either discrete or continuous. Here, we focus on continuous latent variables with a Gaussian marginal 
distribution. We consider both independent and dependent latent variables, and we formalize two different types of dependence.\\


\textbf{A2.} We assume that the latent variable $\mathbf{Z} \in \mathbb{R}^{q}$  has a standard normal distribution, such that $E[\mathbf{Z}'\mathbf{Z}]=\boldsymbol{I}_{q}$. The 
density function is
\beq \label{Eq: Den_z_A}
h(\mathbf{z}) = (2\pi)^{-q/2}\exp\l(-\frac{\mathbf{z}^{'} \boldsymbol{I}_{q} \mathbf{z}}{2}\r).
\eeq

\textbf{A2$^\prime$.} We assume that the latent variable $\mathbf{Z} \in \mathbb{R}^{q}$ has a normal distribution with mean vector zero and covariance matrix $\boldsymbol \Sigma = E[\mathbf{Z}'\mathbf{Z}]$. Thus, the density function becomes
\beq \label{Eq: Den_zS_A}
h(\mathbf{z}) = (2\pi)^{-q/2}\vert \det(\boldsymbol \Sigma)\vert^{-1/2}\exp\l(-\frac{\mathbf{z}^{'} \boldsymbol \Sigma^{-1}\mathbf{z} }{2}\r).
\eeq

We take a conditional independence approach: we assume that the values of the random field $\mathcal{Y}$ are mutually independent given the latent variables $\textbf{Z}$. That is:\\

\textbf{A3.} Edges are assumed to be conditionally independent given the latent variables.\\

Finally, we assume: \\

\textbf{A4.} 
 For each pair $(i,j)\in V\times V$, the random variable $Y_{ij}$ has a distribution belonging to the exponential family, whose canonical parameter can be expressed as a linear combination of latent factors, so the  density  function  is given by:
\begin{equation}
P(y_{ij}\vert \mathbf{z}) = \exp\l\{ [y_{ij} (\boldsymbol{\alpha}_{ij}' \mathbf{z}) - b_{ij}(\boldsymbol{\alpha}_{ij}' \mathbf{z}) ]/\varphi_{ij} + c_{ij}(y_{ij},\varphi_{ij})  \r\}, 
\label{Eq: Pyij}
\end{equation}

\noindent where: $\boldsymbol{\alpha}_{ij}$ are latent weights  determining the effect of the latent variables $\textbf{Z}$. Without loss of generality we can either consider a no-intercept model, where $\boldsymbol{\alpha}_{ij} \in \mathbb{R}^{q}$, or a model with an intercept term and thus $\boldsymbol{\alpha}_{ij} \in \mathbb{R}^{q+1}$. The $b_{ij}$, $c_{ij}$  are known functions depending on the selected member of the exponential family and are standard in the GLM setting;  $\varphi_{ij}$ is the scale parameter. We refer  to
\citet{McN89} for a book length introduction. To illustrate the functional form of these quantities, in Section \ref{Sec: adj}, we give a detailed examples in the case of Bernoulli response and sketch the key equations in the case of Poisson response. 
Extending the terminology of \citet{BKM11}, we label the  model in (\ref{Eq: Pyij}) Graph Generalized Linear Latent Variable Model (GGLLVM). For the sake of exposition, in the rest of the paper we focus on a \texttt{binary} adjacency matrix $\textbf{Y}$. However, we emphasize that the GGLLVMs provide a flexible setting and they can also allow for categorical variables (where for instance each $Y_{ij}$ follows a multinomial distribution) and for continuous variables (where for instance each $Y_{ij}$ follows a Gaussian distribution).

\subsection{Discussion on model assumptions} \label{Sec: discass}
The modeling assumptions \textbf{A1}, \textbf{A2} (or \textbf{A2$^{\prime}$}), and \textbf{A3} are general and various widely-used models satisfy them; we refer to
\citet{RFR16} and reference therein. 

Assumptions \textbf{A2-A2$'$}  express the structure of the latent variables over the network: the $q$-dimensional latent vector $\mathbf{Z}$ impacts on all of the edges through {(\ref{Eq: Pyij}), which states the probability of a node being connected with any other node}. The main difference between \textbf{A2-A2$'$} is that in \textbf{A2} we assume that the latent variables are uncorrelated (actually, they are independent, due to the Gaussian assumption), whilst in \textbf{A2$'$}  we assume that there exists a covariance among them. 
One might think of introducing observable covariates,  which affect each dyad $Y_{ij}$. These dyadic covariates are exogenous and observable factors. For instance,   
as far as import/export modeling is concerned,  the covariates may include the presence of commercial agreements between countries and/or tax deductions for specific items. Similarly, continuous dyadic covariates are also possible. We may interpret the latent factors defined in \textbf{A2} and \textbf{A2$^{\prime}$} as latent or omitted dyadic covariates. If a vector  $\boldsymbol{x}$ of observed dyadic covariates is present, the model for the conditional density function (\ref{Eq: Pyij}) becomes 
\begin{equation}
P(y_{ij}\vert \mathbf{z},\boldsymbol{x}) = \exp\l\{ [y_{ij} (\boldsymbol{\alpha}_{ij}' \mathbf{z} + \boldsymbol{\beta}_{ij}'\boldsymbol{x} ) - b_{ij}(\boldsymbol{\alpha}_{ij}' 
\mathbf{z} + \boldsymbol{\beta}_{ij}'\boldsymbol{x}) ]/\varphi_{ij} +  c_{ij}
(y_{ij},\varphi_{ij})   \r\}, \label{Eq.  PX}
\end{equation}
where $\boldsymbol{x}$ is a $L \times 1$ covariate effecting all the dyads of the network and $\boldsymbol{\beta}_{ij}$ is a $L$-dimensional parameter.
To simplify our exposition, we do not pursue any further this model, i.e. we assume that $\boldsymbol{\beta}_{ij} \equiv \boldsymbol{0}$, for every dyad. However, we stress that our framework naturally permits the inclusion of covariates, under the same code implementation.

Assumption \textbf{A4} is in line with the GLM structure as in 
Section 2 of \citet{HRH02}. With this regard, we remark that, in (\ref{Eq: Pyij}), the vector parameters (including the intercept)  $\boldsymbol{\alpha}_{ij}'\in\mathbb{R}^{q+1}$ for \textbf{A2} and \textbf{A2$^{\prime}$} 
are factor loadings, which express the influence that latent variables have on $Y_{ij}$. We refer to \citet{BKM11}. 

Our model is characterized by the dot product  between two different types of latent variable: $\boldsymbol{\alpha}$ and $\textbf{Z}$. As far as this aspect is concerned, our approach shares similarities with some bipartite network models and with  the latent projection models. To elaborate further,  \citet{Gormley200790}  and \citet{Friel20166629} consider bipartite network models whereby two separate sets of latent variables come into play. This is similar to our modeling, where both $\boldsymbol{\alpha}$ and $\textbf{Z}$ determine the probability in (\ref{Eq: Pyij}) or in (\ref{Eq. PX}). Specifically, the $\boldsymbol{\alpha}$ parameters encode, for each node, the \textit{sender} effects $\boldsymbol{\alpha}_{i\cdot}$ and the \textit{receiver} effects $\boldsymbol{\alpha}_{\cdot j}$: they determine the propensity of node $i$ to send edges, and, respectively, of node $j$ to receive edges, to or from any other node.
The latent factors $\textbf{Z}$ characterize instead the different layers of a multiview network.
This 
formulation permits additional flexibility, since the factors $\boldsymbol{\alpha}$ may differ for each individual edge, which in turn
yields the possibility of representing a variety of relevant network features, such as edge reciprocation and triangles. Our model does not specifically address these particular network motifs, however it is capable of representing them. We refer to the top panel of Figure \ref{Fig_GLAMLE} for a schematic representation. A particular aspect that our model can capture in a very natural way is the persistence of features across different layers. 
This is due to the fact that $\boldsymbol{\alpha}$ is identical across network layers, and so the presence of a given triangle in a specific layer would make that triangle more likely also on all other layers.
Finally, similarly to the projection models of \citet{HRH02}, the interpretation of our model relies on studying whether the points in the latent space tend to align towards the same direction or not. This means that, whenever two latent factors align, we observe a higher predictor, whereas lower connection probabilities are generated by two factors pointing in opposite directions. 
In addition, the distance of either $\boldsymbol{\alpha}$ or $\textbf{Z}$ from the center of the space $(0,0)$ can also be a determining aspect in defining the connectivity of nodes, edge-wise or layer-wise, respectively.
We refer to Section  \ref{Sec: application} for a discussion on these aspects in the context of a real data example.


\section{Methodology: the GLAMLE} \label{Sec: GLAMLE}

Assume we are given  random samples $\boldsymbol Y^{(1)}, \boldsymbol Y^{(2)}, \cdots, \boldsymbol Y^{(K)}$ of a network, representing $K$ different types of relational ties between the $n_V$ actors---{namely, $K$ network views on the same set of nodes}. 

We set up a GLLVM model for $\{\boldsymbol{Y}^{(k)}\}$ and we aim at deriving an estimation procedure for the model parameter {$\bfth=(\text{vec}(\boldsymbol{\alpha}), \text{vec}(\boldsymbol{\varphi}), \boldsymbol{\Sigma})'$, where $\text{vec}(\boldsymbol{\alpha})$ and $\text{vec}(\boldsymbol{\varphi})$ contain all the loadings and the scale parameters, respectively}. In principle, we  should resort to the maximum likelihood estimator (MLE). However, this involves integrating out the latent factors $\textbf{Z}$, a task which typically entails 
a complicated likelihood structure, not tractable analytically. To conduct inference on $\bfth$, extant estimation procedures for GLLVM and/or for graph models rely
on Bayesian methods, which hinges on MCMC estimation procedures. Other approaches in the GLLVM context propose 
the use of variational approximations; see e.g. \citet{HWOH17}. Here, we propose to use a Laplace approximation as in \citet{HRVF04} and we explain how this method can be adapted to our GGLLVM. 
We illustrate in detail how to derive the GLAMLE in the case of the adjacency matrix (binary dyadic variables). In Appendix \ref{Sec: uniq} we explain how to ensure uniqueness of the solution to the estimating equations that define the GLAMLE. In Appendix \ref{Section: AppB}, we provide
the estimating equations for the Laplace-approximated likelihood in the case of count data (namely, Poisson dyadic variables). Thanks to the flexibility of the GLLVM, other relational data (see Section \ref{Sec: concl}) can be considered: the key methodological steps for the derivation of the corresponding GLAMLE remain the same as the ones described in this section.

\subsection{Adjacency matrix (binary random variables)} \label{Sec: adj}

Let $Y_{ij}\vert \mathbf{Z}$ be independent Bernoulli distributions with mean $\pi_{ij}$ (respectively, for all $i$ and $j$), where $\mathbf{Z}=(1,Z_1,\cdots, Z_q)'=(1,\mathbf{Z}'_{(2)})' \in \mathbb{R}^{q+1}$, for $\mathbf{Z}'_{(2)} \in \mathbb{R}^q$, where $q \ll n_V$ and $n_V,q \in \mathbb{N}$. Then, $Y_{ij}$ have conditional expectation equal to $\pi_{ij}$, and using the canonical link function we have
\beq \label{Eq: pi}
P(Y_{ij}=1\vert \mathbf{Z}=\mathbf{z}) = 1- P(Y_{ij}=0 \vert \mathbf{Z}=\mathbf{z}) = \pi_{ij}= \frac{\exp( \boldsymbol{\alpha}'_{ij}\mathbf{z})}{1+\exp( \boldsymbol{\alpha}'_{ij}\mathbf{z})}. 
\eeq
Thus, (\ref{Eq: Pyij}) becomes 
$g_{ij}(y_{ij}\vert \mathbf{z}) = \exp\l\{y_{ij}\boldsymbol{\alpha}'_{ij}\mathbf{z}-\log\l[1+\exp\l( \boldsymbol{\alpha}'_{ij}\mathbf{z}\r)\r]\r\}.$ 
{The function $g_{ij}$ depends on the parameter vector $\boldsymbol{\alpha}'_{ij}$, since the scale parameter of \eqref{Eq: Pyij} is set to $1$ and  $\bfth=\text{vec}(\boldsymbol{\alpha})$. However, for ease-of-notation, we remove the dependence on $\boldsymbol{\alpha}'_{ij}$ from here onwards.}

The complete data likelihood for $\boldsymbol y=\{y_{ij} : i,j \in (V \times V)\}$ 
is $\prod_{i\ne j }^{n_V}g_{ij}(y_{ij}\vert \mathbf{z})h(\mathbf{z}_{(2)})$,
where $h(\mathbf{z}_{(2)})$ is the density function of latent variables.  As it is common in the literature on network models, see e.g.  \citet{WW87} or \citet{KC14}, it is assumed that there are no self edges, so the diagonal entries of $\boldsymbol Y$ are all structural zeros. Integrating out the latent variables, we get the marginal density function
\beq \label{Eq: MarDen}
f_{\boldsymbol{\alpha}}(\boldsymbol y)= \int_{} \l\{ \prod_{i\ne j }^{n_V}g_{ij}(y_{ij}\vert \mathbf{z})\r\}h(\mathbf{z}_{(2)}) d\mathbf{z}_{(2)},
\eeq
where $\boldsymbol{\alpha}$ is the $m\times (q+1)$ matrix containing all the $\boldsymbol{\alpha}_{ij}$.
Under \textbf{A2} and \textbf{A2$^{\prime}$} the size of the integrals $q$ is fixed and does not grow with the sample size. This is a pivotal computational aspect of our approach: we are in a setting similar to the one described by \citet{V96} for non linear mixed effects models, where the dimension of the integral  to be approximated by the Laplace method is $q$ (namely it is the dimension of $\mathbf{z}_{(2)}$) and it does not increase with the sample size; see section \ref{Sec: BA2} and \ref{Sec: BA2p}. 
Given  $K$ random samples $\boldsymbol Y^{(1)}, \boldsymbol Y^{(2)}, \cdots, \boldsymbol Y^{(K)}$, 
the log-likelihood is 
%

\beq \label{Eq: ell}
\ell(\boldsymbol{\alpha}) =\sum_{k=1}^{K}\log\l(\int_{} \l\{ \prod_{i\ne j }^{n_V}g_{ij}(Y^{(k)}_{ij}\vert \mathbf{z})\r\}h(\mathbf{z}_{(2)}) d\mathbf{z}_{(2)}\r),
\eeq
where the random variable $Y^{(k)}_{ij}$ represents the response for the edge $(i,j)\in \{V \times V\}$ and $\mathbf{z}$ represents $q$ latent variables. The total sample size is $n=n^2_V K$. 
Next we explain how to derive an approximate MLE using a Laplace approximation to the untreatable likelihood in (\ref{Eq: ell}). We perform the derivation, under the 
assumptions (\textbf{A2})-(\textbf{A2
$^{\prime}$}), namely considering different dependence structures for the latent variables. 
%

\subsubsection{Uncorrelated latent variables (A2)} \label{Sec: BA2}

Let us assume that the latent variables $\mathbf{z}_{(2)}$ have standard normal distributions and that they are independent. The corresponding density is
\beq \label{Eq: Den_z}
h(\mathbf{z}_{(2)}) = (2\pi)^{-q/2}\exp\l(-\frac{\mathbf{z}^{'}_{(2)}\mathbf{z}_{(2)}}{2}\r).
\eeq
Then, we rewrite the marginal density function as the following expression, 
\beq
f_{\boldsymbol{\alpha}}(\boldsymbol y^{(k)})= \int \exp\l\{ m Q\l(\boldsymbol{\alpha},\mathbf{z},\boldsymbol y^{(k)}\r)\r\} d\mathbf{z}_{(2)},
\eeq
with the function
\beq \label{Eq: Q}
Q\l(\boldsymbol{\alpha},\mathbf{z},\boldsymbol y^{(k)}\r)= m^{-1}\l[  \sum_{i\ne j}^{n_V}\l\{y_{ij}^{(k)}\boldsymbol{\alpha}'_{ij} \mathbf{z}- \log\l(1+\exp\l(\boldsymbol{\alpha}'_{ij}\mathbf{z}\r)\r)\r\}-\frac{\mathbf{z}'_{(2)}\mathbf{z}_{(2)}}{2}-\frac{q}{2}\log(2\pi)\r],
\eeq
where  $\sum_{i \ne j}^{n_V} = \sum_{i=1}^{n_V}\sum_{j=i+1}^{n_V}$ 
for undirected relations, and 
$\sum_{i \ne j}^{n_V} = \underset{i \ne j}{\sum_{i=1}^{n_V}\sum_{j=1}^{n_V}}$
for directed relations.
We derive the Laplace-approximated marginal density function
$$ 
\tilde{f}_{\boldsymbol{\alpha}}(\boldsymbol{y}^{(k)}) = \l(\frac{2\pi}{m}\r)^{q/2}\det\l\{ -U\l(\hat{\mathbf{z}}^{(k)}\r)\r\}^{-1/2}\exp\l\{mQ\l(\boldsymbol{\alpha},\hat{\mathbf{z}}^{(k)},\boldsymbol{y}^{(k)}\r)\r\},
$$
where 
\beq
U\l(\boldsymbol{\alpha},\hat{\mathbf{z}}^{(k)}\r)=\left. \frac{\partial^2 Q\l(\boldsymbol{\alpha},\mathbf{z},\boldsymbol y^{(k)}\r)}{\partial \mathbf{z}'\partial \mathbf{z}}\right \vert_{\mathbf{z}=\hat{\mathbf{z}}^{(k)}}=-m^{-1}\Gamma\l(\boldsymbol{\alpha},\hat{\mathbf{z}}^{(k)}\r),
\eeq   
\beq
\Gamma\l(\boldsymbol{\alpha},\hat{\mathbf{z}}^{(k)}\r)=\sum_{i \ne j}^{n_V} \l\{\frac{\exp\l(\boldsymbol{\alpha'}_{ij}\hat{\mathbf{z}}^{(k)}\r)\boldsymbol{\alpha}_{ij(2)}\boldsymbol{\alpha}'_{ij(2)} }{\l(1+\exp\l(\boldsymbol{\alpha}'_{ij}\hat{\mathbf{z}}^{(k)}\r)\r)^2}\r\} + \boldsymbol{I}_q,
\eeq                          
with $\boldsymbol{\alpha}_{ij}=(\alpha_{ij,0}, \boldsymbol{\alpha}_{ij(2)}')'$ and $\hat{\mathbf{z}}^{(k)}=(1, (\hat{\mathbf{z}}_{(2)}^{(k)})')'$ maximizing $Q\l(\boldsymbol{\alpha},\mathbf{z},\boldsymbol y^{(k)}\r)$, therefore being the solution to  
$\partial Q\l(\boldsymbol{\alpha},\mathbf{z},\boldsymbol{y}^{(k)}\r)/\partial \mathbf{z}=\mathbf 0$ and it is such that
\beq \label{Eq: Est_z}
\hat{\mathbf{z}}_{(2)}^{(k)}=\hat{\mathbf{z}}_{(2)}^{(k)}\l(\boldsymbol{\alpha},\boldsymbol{y}^{(k)}\r)=\sum_{i \ne j}^{n_V} \l\{y_{ij}^{(k)}-\frac{\exp\l(\boldsymbol{\alpha}'_{ij}\hat{\mathbf{z}}^{(k)}\r)}{1+\exp\l(\boldsymbol{\alpha}'_{ij}\hat{\mathbf{z}}^{(k)}\r)}\r\}\boldsymbol{\alpha}_{ij(2)}.   \quad \quad  
\eeq

\begin{figure}[htp!] 
\begin{center}
\includegraphics[width=0.6\textwidth, height=0.25\textheight]{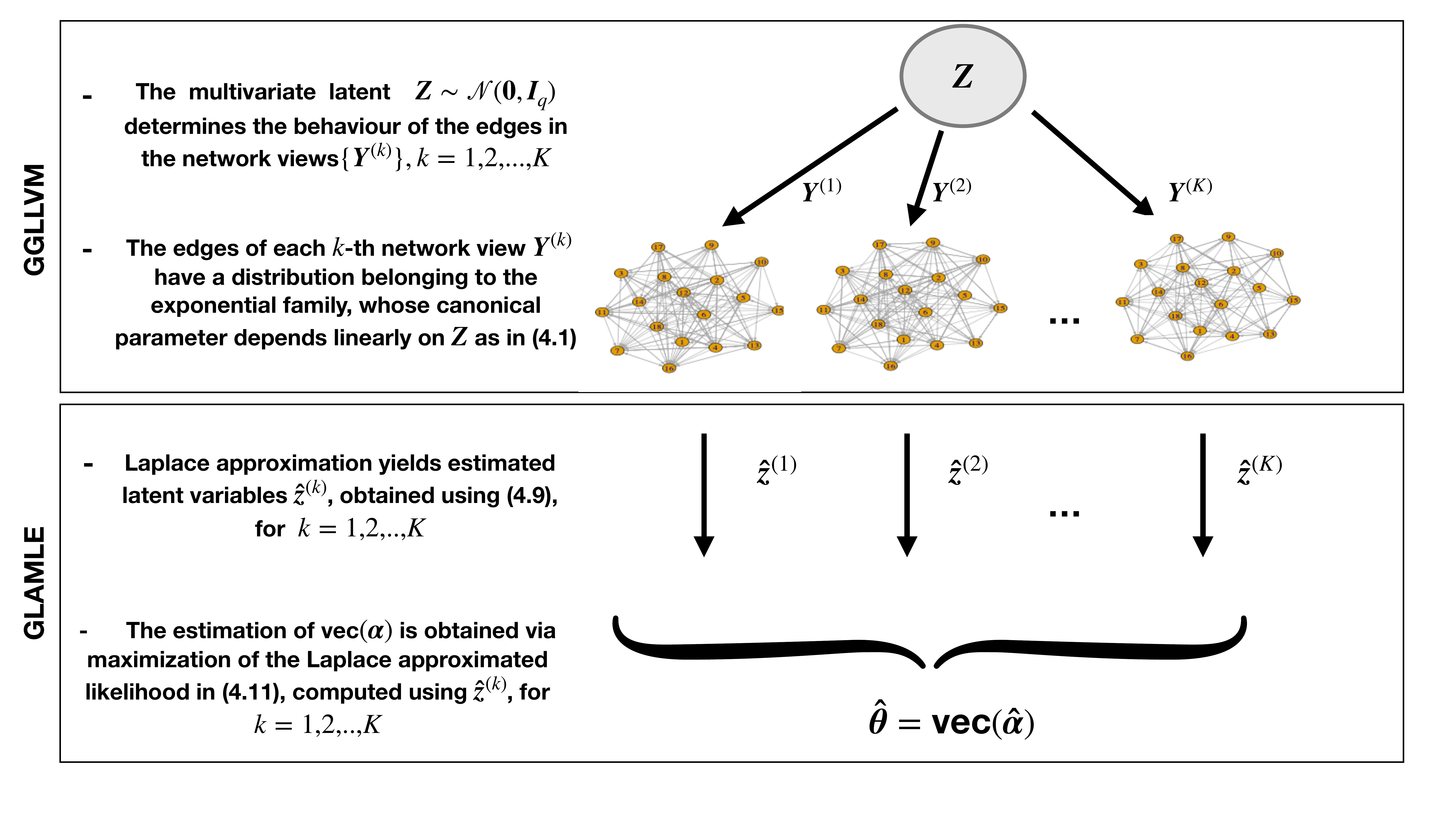} 
\end{center}
\caption{Schematic illustration of the data generating mechanism (top panel) and of the Graph Laplace Approximated estimation method (bottom panel), in the case of Bernoulli  variables (adjacency matrix entries), under Assumption A2.}
\label{Fig_GLAMLE}
\end{figure}

Eq. (\ref{Eq: Est_z}) deserves some comments. In the Laplace approximation, the latent factors $\mathbf{z}$ are implicitly treated as parameters, which are needed to approximate the integrals characterizing the likelihood. Therefore, $\hat{\mathbf{z}}_{(2)}^{(k)}$ can be formally interpreted as the maximum likelihood estimates of the latent factors in the $k$-th network view: each $\hat{\mathbf{z}}_{(2)}^{(k)}$ depends on the model parameter $\bfth$ and on the observation $\boldsymbol{y}^{(k)}$. Moreover, thanks to the Laplace approximation, we have
\begin{eqnarray}
f_{\boldsymbol{\alpha}}(\boldsymbol{y}^{(k)}) &=& \tilde{f}_{\boldsymbol{\alpha}}(\boldsymbol{y}^{(k)})\l\{1+O\l(m^{-1}\r)\r\}. \label{Eq. Lapl}
\end{eqnarray}
Now, for $K$ multiple views of the network, $\boldsymbol Y^{(1)}, \boldsymbol Y^{(2)}, \cdots, \boldsymbol Y^{(K)}$,  the approximate log-likelihood  is
\color{black}
\begin{small}  
\begin{equation} \label{Eq: ellLap}
\begin{split}
 &\tilde{\ell}(\boldsymbol{\alpha})=\sum_{k=1}^{K} \l(-\frac{1}{2}\log\l[\det\l\{\Gamma\l(\boldsymbol{\alpha},\hat{\mathbf{z}}^{(k)}  \r)\r\}\r]\r. + \\
 &\hspace{2cm}\l. +\sum_{i \ne j}^{n_V} \l\{Y_{ij}^{(k)}\boldsymbol{\alpha}'_{ij} \hat{\mathbf{z}}^{(k)}- \log\l(1+\exp\l(\boldsymbol{\alpha}'_{ij}\hat{\mathbf{z}}^{(k)}\r)\r)\r\}-\frac{\l(\hat{\mathbf{z}}^{(k)}_{(2)}\r)'\hat{\mathbf{z}}^{(k)}_{(2)}}{2}\r).  \\
\end{split}
\end{equation}
\end{small}  
The parameter $\boldsymbol{\alpha}$ is obtained by solving the equation where we set the derivative of the approximate log-likelihood equal to the zero vector:
\begin{small}  
\begin{equation} \label{Eq: Partial_Alpha}
\frac{\partial \tilde{\ell}(\boldsymbol{\alpha})}{\partial \alpha_{ij,l}} = \sum_{k=1}^{K} \l(-\frac{1}{2}\tr\l\{\Gamma\l(\boldsymbol{\alpha},\hat{\mathbf{z}}^{(k)}\r)^{-1}\frac{\partial \Gamma\l(\boldsymbol{\alpha},\hat{\mathbf{z}}^{(k)}\r)}{\partial \alpha_{ij,l}}  \r\}+ \sum_{i \ne j}^{n_V} \l\{Y_{ij}^{(k)} - \frac{\exp\l(\boldsymbol{\alpha}'_{ij}\hat{\mathbf{z}}^{(k)}\r)}{1+\exp\l(\boldsymbol{\alpha}'_{ij}\hat{\mathbf{z}}^{(k)}\r)}\r\}\hat{{z}}^{(k)}_{l}\r)= 0,
\end{equation}
\end{small}
where $\alpha_{ij,l}$ and $\hat{{z}}^{(k)}_{l}$ are the $l_{th}$ element of the vectors $\boldsymbol{\alpha}_{ij}$ and $\hat{\mathbf{z}}^{(k)}$ respectively. We label the resulting M-estimator, the Graph Laplace-approximated maximum Likelihood estimator, in short GLAMLE.
To summarize the GGLLVM modeling and the GLAMLE inference, in Figure \ref{Fig_GLAMLE} we display a schematic representation of the GGLLVM data generating mechanism (based on Assumptions  \textbf{A1}-\textbf{A4} and with uncorrelated latent variables) and the described inference approach for the $\text{vec}(\boldsymbol{\hat{\alpha}})$.

\subsubsection{Correlated latent variables (A2$^{\prime}$)} \label{Sec: BA2p}

We have correlated latent variables with a correlation matrix $\boldsymbol \Sigma$ such that its density function becomes
\beq \label{Eq: Den_z_A2}
h(\mathbf{z}_{(2)}) = (2\pi)^{-q/2}\vert \det(\boldsymbol \Sigma)\vert^{-1/2}\exp\l(-\frac{\mathbf{z}^{'}_{(2)} \boldsymbol \Sigma^{-1}\mathbf{z}_{(2)}}{2}\r).
\eeq

Then, the function $Q$ as in (\ref{Eq: Q}) is modified as follows
\bea \label{Eq: Q_Corr}
Q\l(\boldsymbol{\alpha},\mathbf{z},\boldsymbol{y}^{(k)}, \boldsymbol \Sigma\r)&=& \frac{1}{m}\l[  \sum_{i\ne j}^{n_V}\l\{y_{ij}^{(k)}\boldsymbol{\alpha}'_{ij} \mathbf{z}- \log\l(1+\exp\l(\boldsymbol{\alpha}'_{ij}\mathbf{z}\r)\r)\r\} \right.  \nn \\
&-& \left. \frac{\mathbf{z}'_{(2)}\boldsymbol \Sigma^{-1}\mathbf{z}_{(2)}}{2}-\frac{1}{2}\log|\det(\boldsymbol \Sigma)|-\frac{q}{2}\log(2\pi)\r].
\eea
Thus, the iterative equation (\ref{Eq: Est_z}) becomes
\beq \label{Eq: Est_z_Corr}
\hat{\mathbf{z}}_{(2)}^{(k)}=\hat{\mathbf{z}}_{(2)}^{(k)}\l(\boldsymbol{\alpha},\boldsymbol{y}^{(k)},\boldsymbol\Sigma \r)=\sum_{i \ne j}^{n_V} \l\{y_{ij}^{(k)}-\frac{\exp\l(\boldsymbol{\alpha}'_{ij}\hat{\mathbf{z}}^{(k)}\r)}{1+\exp\l(\boldsymbol{\alpha}'_{ij}\hat{\mathbf{z}}^{(k)}\r)}\r\}\boldsymbol\Sigma\boldsymbol{\alpha}_{ij(2)},   \quad \quad  k = 1,\cdots, n.
\eeq
We resort to the Laplace approximation as in (\ref{Eq. Lapl}). Then, for a random sample $\boldsymbol Y^{(1)}, \boldsymbol Y^{(2)}, \cdots, \boldsymbol Y^{(K)}$, the Laplace approximate log-likelihood becomes
\bea  \label{Eq: ellLap}
\tilde{\ell}(\boldsymbol{\alpha}, \boldsymbol \Sigma)&=&\sum_{k=1}^{K} \l(-\frac{1}{2}\log\l[\det\l\{\Gamma\l(\boldsymbol{\alpha},\hat{\mathbf{z}}^{(k)} , \boldsymbol \Sigma \r)\r\}\r] -\frac{1}{2}\log|\det(\boldsymbol \Sigma)|  \r)\\ \nn
&+&\sum_{k=1}^{K} \l( \sum_{i \ne j}^{n_V} \l\{Y_{ij}^{(k)}\boldsymbol{\alpha}'_{ij} \hat{\mathbf{z}}^{(k)}- \log\l(1+\exp\l(\boldsymbol{\alpha}'_{ij}\hat{\mathbf{z}}^{(k)}\r)\r)\r\}-\frac{\l(\hat{\mathbf{z}}^{(k)}_{(2)}\r)'\boldsymbol \Sigma^{-1}\hat{\mathbf{z}}^{(k)}_{(2)}}{2}\r),
\eea
where 
\beq
\Gamma\l(\boldsymbol{\alpha},\hat{\mathbf{z}}^{(k)}, \boldsymbol \Sigma \r)=\sum_{i \ne j}^{n_V} \l\{\frac{\exp\l(\boldsymbol{\alpha}'_{ij}\hat{\mathbf{z}}^{(k)}\r)\boldsymbol{\alpha}_{ij(2)}\boldsymbol{\alpha}'_{ij(2)} }{\l(1+\exp\l(\boldsymbol{\alpha}'_{ij}\hat{\mathbf{z}}^{(k)}\r)\r)^2}\r\} + \boldsymbol\Sigma^{-1}.
\eeq 

The estimated $\boldsymbol{\alpha}$ is the root of (\ref{Eq: Partial_Alpha}) and we notice that, in this case, $\Gamma\l(\boldsymbol{\alpha},\hat{\mathbf{z}}^{(k)}, \boldsymbol \Sigma \r)$ depends on $\boldsymbol \Sigma $. To estimate the $\boldsymbol \Sigma$ matrix, we resort on the derivative of the approximate log-likelihood with respect to $\sigma_{ij,l}$, which yields
\begin{eqnarray} 
\frac{\partial \tilde{\ell}(\boldsymbol{\alpha},\boldsymbol{\Sigma})}{\partial \sigma_{ij,l}} &=& \sum_{k=1}^{K} \l(-\frac{1}{2}\tr \l
\{\Gamma\l(\boldsymbol{\alpha},\hat{\mathbf{z}}^{(k)}, \boldsymbol \Sigma \r)^{-1}\boldsymbol \Sigma^{-1}\frac{\partial\boldsymbol \Sigma}{\partial 
\sigma_{ij,l}}\boldsymbol\Sigma^{-1}  \r\} \r) \nn \\
&+&\sum_{k=1}^{K} \l( - \frac{1}{2} \tr \l(\boldsymbol{\Sigma}^{-1}\frac{\partial \boldsymbol{\Sigma}}{\partial \sigma_{ij,l}} \r)
+ \frac{1}{2} \l (\hat{\mathbf{z}}^{(k)}_{(2)}\r)'\boldsymbol{\Sigma}^{-1}\frac{\partial \boldsymbol{\Sigma}}{\partial \sigma_{ij,l}}
\boldsymbol{\Sigma}^{-1}\hat{\mathbf{z}}^{(k)}_{(2)}\r). \label{Eq: Partial_Sigma}
\end{eqnarray}

\section{Statistical aspects } \label{Sec: Infe}

The GLAMLE $\boldsymbol{\hat{\bfth}}$ is  an M-estimator, which  is implied by setting to zero the equations derived from the first order conditions of the Laplace-approximated likelihood; see among the others \citet{H81} and  \citet{vdW98} for book-length introduction to M-estimation, while see \citet{HRVF04} for a discussion in the setting of Laplace approximation. To elaborate further, let us define $\boldsymbol{\tilde{\mathcal S}}(\bfth)= {\partial \tilde{\ell}} / {\partial{\bfth}}$, whose $k$-th component is denoted by 
$\boldsymbol{\tilde{\mathcal S}}^{(k)}({\bfth}) = \nabla_{\bfth} \ln \tilde{f}_{\boldsymbol{\bfth}}(\boldsymbol{y}^{(k)})$.  By definition,  ${\hat{\bfth}}$ 
solves 
\begin{equation}
\boldsymbol{\tilde{\mathcal S}}(\boldsymbol{{\bfth}})= \sum_{k=1}^{K} \boldsymbol{\tilde{\mathcal S}}^{(k)}({{\bfth}}) = \boldsymbol{0}. \label{Eq. GLAMLEshort}
\end{equation}
 In this section, we discuss various statistical aspects (like e.g.\ uniqueness, existence, asymptotic distribution) related to $\hat\bfth$.


\subsection{Interpretation} \label{Sec: Inter}

 Let $g_{ij}(y^{(k)}_{ij}\vert \mathbf{z})$ be any distribution belonging to the exponential family (not necessary a Binomial or a Poisson) indexed by $\boldsymbol\theta$ and for which a GGLLVM can be written and such that the likelihood is 
$$ 
\ell(\boldsymbol{\boldsymbol\theta}) =\sum_{k=1}^{K}\log\l(\int_{} \l\{ \prod_{i\ne j }^{n_V}g_{ij}(Y^{(k)}_{ij}\vert \mathbf{z})\r\}h(\mathbf{z}_{(2)}) d\mathbf{z}_{(2)}\r).
$$
Denoting as $\text{int}(\boldsymbol{\Theta})$ the interior of 
the parameter space $\boldsymbol{\Theta}$, we remark that according to \textbf{A1}-\textbf{A4}, the class of densities $\{f_{\bfth}\}$, for $\bfth \in \boldsymbol{\Theta}$,  contains the 
actual density $f_0$, thus there exists $\boldsymbol{\theta}_0 \in \text{int}\boldsymbol{\Theta}$ such that $f_0 \equiv f_{\bfth_{0}}$.
The existence of $\boldsymbol{\theta}_0$ follows from Theorem 2.1 in \citet{W82}.

Now, let us 
define $\bfS = {\partial {\ell}} / {\partial{\bfth}}$. In analogy with (\ref{Eq. GLAMLEshort}), 
the contribution of the $k$-th sample to the exact likelihood score is $\boldsymbol{\mathcal{S}}^{(k)}({\bfth})=\nabla_{\bfth} \ln {f}_{\boldsymbol{\bfth}}(\boldsymbol{y}^{(k)})$, so $\bfS = \sum_{k=1}^{K} \boldsymbol{\mathcal{S}}^{(k)}({\bfth}) $.
Under $\bfth_0$ the  Fisher consistency of the MLE holds for any $k$, thus (occasionally, we will write $E_{0}$ in lieu of $E_{f_0}$)
$
E_{0}[\boldsymbol{\mathcal{S}}^{(k)}({\bfth_{0}})] = \boldsymbol{0}  \Rightarrow 
E_{0}[\boldsymbol{\mathcal{S}}({\bfth_{0}})] = \boldsymbol{0}.
$


Due to its approximate nature, the function $\tilde\ell$ is a pseudo likelihood, in the sense of \citet{W82}. 
Indeed, 
we may interpret the class of Laplace-approximated densities 
$\{\tilde{f}_{\bfth}\}$ as a collection of
misspecified densities. Thus $\partial \tilde \ell /\partial \boldsymbol\theta = \boldsymbol{0}$ defines a system of estimating equations whose solution is a maximum 
likelihood estimator obtained using a misspecified likelihood; sometimes this is called a Quasi MLE (QMLE),  see e.g. \citet{W96} or \citet{A16} for book-length 
presentations. In the presence of misspecification, the asymptotic covariance matrix of the GLAMLE no longer equals the inverse of Fisher's information matrix. 
Nevertheless, the covariance matrix can be consistently estimated and it simplifies to the familiar form (that is the expected value of the Hessian of the log-likelihood) 
in the absence of misspecification. More in detail, we define the Kullback-Leibler Information Criterion (KLIC) as
\beq
I(f_{\bfth_0}: \tilde{f}_{\bfth}) = E_{0}  \left[  \ln \left( \frac{f_{\bfth_0}(\boldsymbol U)}{\tilde{f}_{\bfth}(\boldsymbol U)}    \right )     \right]  \equiv I(f_{0}: \tilde{f}_{\bfth}) = E_{0}  \left[  \ln \left( \frac{f_{0}(\boldsymbol U)}{\tilde{f}_{\bfth}(\boldsymbol U)}    \right )     \right], 
\label{Eq. KLIC}
\eeq
and 
we define $\tilde{\bfth} = \arg\min_{\bfth \in \boldsymbol{\Theta}} I(f_{\bfth_0}: \tilde{f}_{\bfth})$ and label it as \textit{pseudo-true value}. By the above construction, the pseudo-value is such that we are minimizing our ignorance about the true structure. {In words, we are considering the pseudo parameter value which minimizes the KLIC  between the misspecified density and the true one.} Therefore,  the resulting QMLE  can be interpreted as a minimum ignorance estimator. To the best of our knowledge, this interpretation of the Laplace-approximated MLE is new on the literature about GLLVM; see  \citet{HRVF04}, \citet{HuSVF09} and related papers.

\subsection{Asymptotics}

With the interpretation of Section \ref{Sec: Inter} in mind,  we study the  behavior of the GLAMLE under two different asymptotic schemes. First, in section \ref{Sec: Asy1}, we consider the case where $n_V$ is fixed, while $K$ diverges toward infinity. We call it \textit{asymptotic regime (i)}: the GLAMLE is a misspecified MLE and we consider its asymptotic distribution, studying its consistency to the pseudo-true value.
Then,  in section \ref{Sec: bias}, we consider  the behavior of the GLAMLE when $m = O(K^{\varrho})$, for some  $\varrho > 0$, and $K$ diverges to infinity. We call it \textit{asymptotic regime (ii)}: we  study  the GLAMLE convergence rate to the true parameter value.

\subsubsection{Asymptotic regime (i): rate of convergence to the pseudo-true value}\label{Sec: Asy1}

In what follows, we fix $n_V$ (so the dimension of $\bfth$ is fix too)  and let $K$ diverge to infinity. {Roughly speaking, in this asymptotic regime, we have infinite views of a finite number of network nodes}.
In our setting, both 
$f_{\bfth_0}$ and $\tilde{f}_{\bfth}$ are Radom-Nykodym derivatives of the corresponding complete data distribution function: Theorem 2.1 in \citet{W82} implies
the existence of $\hat{\bfth}$ solution to (\ref{Eq. GLAMLEshort}), while identifiability constraints can be considered to guarantee its uniqueness. Furthermore, 
we introduce \\

\textbf{A5.} $E_{0}  \left[  \ln \tilde{f}_{\bfth}(\boldsymbol U)      \right] $ exists and $ \vert \ln \tilde{f}_{\bfth}(\boldsymbol  u)    \vert \leq \tilde{m}(\boldsymbol  u)$ , with $m$ being 
integrable  w.r.t. $f_{\bfth_0}$. Moreover, $I(f_{\bfth_0}: \tilde{f}_{\bfth})$ has a unique minimum at $\tilde{\bfth}$. \\

This assumption is loosely related to the classical treatment of maximum likelihood given in \citet{LC53} and it is standard in the asymptotic analysis of (Q)MLE under misspecification. 
In particular, thanks to \textbf{A5}, a weak law of large number applies and  $\boldsymbol{\tilde{\mathcal S}}(\boldsymbol{\hat{\bfth}})/K$ is a natural estimator of 
 $E_{\bfth_0}  \left[  \ln \tilde{f}_{\bfth}(\boldsymbol U)      \right] $. Then, $\hat{\bfth}$ is the MLE in the presence of misspecification. Theorem 2.2 in \citet{W82} implies that $\hat{\bfth}$ is a consistent estimator of the pseudo-true (see \citet{GM95}, Ch. 8) value $\tilde{\bfth}$, namely $\hat{\bfth}\overset{\mathcal{P}}{\rightarrow}\tilde{\bfth}$.

Beside addressing the convergence rate, the theory of misspecified maximum likelihood allows us to identify the asymptotic distribution of $\hat{\bfth}$, under the asymptotic regime (i). Indeed, 
assuming the same conditions as in \citet{H81}, pages 131-133, and using standard M-estimation asymptotic theory, we have
\begin{equation}
K^{1/2} (\hat{\bfth}- \tilde{\bfth})\overset{\mathcal{D}}{\rightarrow}\mathcal{N}(0, \boldsymbol V(\tilde{\bfth})),
\label{Asym1}
\end{equation}
where the matrices are
$
\boldsymbol V(\tilde{\bfth}) = \boldsymbol B(\tilde{\bfth})^{-1} \boldsymbol A(\tilde{\bfth})  [\boldsymbol B(\tilde{\bfth})^{-1}]',
$
$
\boldsymbol A(\tilde{\bfth})=E_{0}\left[\left( \nabla_{\bfth} \ln \tilde{f}_{\boldsymbol{\bfth}}(\boldsymbol{y})\vert_{\bfth=\tilde{\bfth}}\right)\left(\nabla_{\bfth'} \ln \tilde{f}_{\boldsymbol{\bfth}}(\boldsymbol{y})\vert_{\bfth=\tilde{\bfth}}\right)  \right], 
$
and 
$
\boldsymbol B(\tilde{\bfth})= -E_{0}\left[. \nabla_{\bfth'} (\nabla_{\bfth} \ln \tilde{f}_{\boldsymbol{\bfth}}(\boldsymbol{y}))\vert_{\bfth=\tilde{\bfth}}   \right].
$
Defining the finite sample counterpart of the above asymptotic variance matrix 
$
\hat{\boldsymbol {V}}({\bfth}) = \hat{\boldsymbol {B}}({\bfth})^{-1} \hat{\boldsymbol {A}}({\bfth}) (\hat{\boldsymbol{B}}({\bfth})^{-1})',
$
with $
\hat{\boldsymbol {A}}({\bfth})= \sum_{k=1}^K K^{-1} \left[\boldsymbol{\mathcal{S}}^{(k)}({\bfth}) \cdot (\boldsymbol{\mathcal{S}}^{(k)}({\bfth}))^T
\right],$ and $
\hat{\boldsymbol {B}}({\bfth})= - \sum_{k=1}^K K^{-1} \nabla_{\bfth^T}  \boldsymbol{\mathcal{S}}^{(k)}({\bfth}),
$
 we have that (element-wise)
 \begin{equation}
 \hat{\boldsymbol{V}}(\hat{\bfth}) \overset{\text{a.s.}}{\rightarrow} \boldsymbol V(\tilde{\bfth}).
 \label{Asym2}
 \end{equation}

From a theoretical standpoint, we remark that due to the misspecification of the approximated likelihood, the GLAMLE $\hat{\bfth}$ does not necessarily converge to $\bfth_0$.  Moreover, the misspecification entails that the second-Bartlett identity does not necessarily hold, namely $\boldsymbol B(\tilde{\bfth})^{-1} \neq \boldsymbol A(\tilde{\bfth})$  (again, to be interpreted element-wise). If this is the case, $\boldsymbol V(\tilde{\bfth})$ does not coincide with the inverse of Fisher information matrix computed at $\tilde{\bfth}$. Nevertheless,  the asymptotic results in (\ref{Asym1})  and (\ref{Asym2}) are useful to construct asymptotic confidence intervals for the pseudo-true value and for hypothesis testing.  For instance, assume that we want  to test $\mathcal{H}_0: h(\tilde{\bfth}) = \boldsymbol 0$ versus $\mathcal{H}_1 : h(\tilde{\bfth}) \neq \boldsymbol 0$, where $h : \boldsymbol \Theta \to \mathbb{R}^r$ is a continuous vector function of ${\bfth}$ such that its Jacobian at the pseudo-true value is finite with full row rank $r$. Then, under $\mathcal{H}_0$, we have that 
$$
\mathcal{W}(\tilde{\bfth}) = K h(\hat{\bfth})' \l[ \nabla_{\bfth} h(\hat{\bfth}) \hat{\boldsymbol {V}}(\hat{\bfth}) \nabla_{\bfth'} h(\hat{\bfth})  \r]^{-1} h(\hat{\bfth}) \sim \chi^2_r,
$$
where $\chi^2_r$ is a central chi-square with $r$ degrees-of-freedom. The statistic $\mathcal{W}$ can be applied to test if some parameters in the Laplace approximated likelihood are equal to zero, e.g. for latent and/or manifest variables selection. Additionally, approximated log-likelihood ratio and/or Lagrange-Multiplier test statistics can be defined, even in the presence of nuisance parameters;  see \citet{W82} and \citet{HR94}.

From a practical standpoint, in our numerical experience, the misspecification bias can be small (if not negligible, see Figure \ref{Fig: RMSE}). This is yielded by the high numerical accuracy of the Laplace approximation (see \ref{Eq. Lapl}). We refer to section \ref{Sec: MC_gllvm} for numerical evidence on this claim.


\subsubsection{Asymptotic regime (ii): rate of convergence to the true value} \label{Sec: bias}



Intuitively, we
expect that either or both $K$ and $m$ need to be sufficiently large for the Laplace approximation to the MLE to work, implying  consistency for the GLAMLE to the true parameter value $\bfth_0$. 
In this subsection, we confirm this intuition by heuristic arguments, in which we make use of formal expansions. 
 A thorough analysis  deserves a separate paper.

To understand how the GLAMLE behaves in the asymptotic regime (ii), we proceed similarly to \citet{V96} and \citet{RVL09}. 
Starting from (\ref{Eq. Lapl}), we write
\begin{equation}
\boldsymbol{\mathcal S}(\hat{\bfth}) = \sum_{k=1}^{K} \l \{   \boldsymbol{\tilde{\mathcal S}}^{(k)}(\boldsymbol{\hat{\bfth}}) 
+ O\l( m ^{-1}  \r)  \r\}, \label{Eq. Sapprox}
\end{equation}
which is similar to Eq. (5) in \citet{V96}. Now, assume that $m = O(K^{\varrho})$, for $\varrho >0$, and define $m_K = O(K^{\varrho_1})$, 
with $\varrho_1 = 1 + \varrho$. Then, we have 
\begin{equation}
K^{-1}\boldsymbol{\mathcal S}(\hat{\bfth}) = K^{-1}  \boldsymbol{\tilde{\mathcal S}}({\hat{\bfth}}) + O_P\l( m_K^{-1}  \r) =  O_P\l( m_K^{-1}  \r),
\label{Eq. Shat}
\end{equation}
where the last equality follows from (\ref{Eq. GLAMLEshort}). A Taylor expansion of the exact likelihood $\bfS$  about ${\bfth}_0$ yields 
$
\boldsymbol{\mathcal S}(\hat{\bfth}) = \boldsymbol{\mathcal S}({\bfth}_0) + 
\mathcal{\mathbf{H}}(\bfth^{\ast}) ({\hat{\bfth}}-{{\bfth}_0}),
$
with $\bfth^{\ast}$ belonging to the segment joining ${\hat{\bfth}}$ and ${{\bfth}_0}$ and the Hessian matrix 
$\mathcal{\mathbf{H}}(\bfth) = \sum_{k=1}^{K} \mathcal{\mathbf{H}}^{(k)}(\bfth)$, where
$\mathcal{\mathbf{H}}^{(k)}(\bfth):={\nabla_{\bfth} \boldsymbol{\mathcal{S}}^{(k)}({\bfth})}$. Then, we have
$
K^{-1} \boldsymbol{\mathcal S}(\hat{\bfth}) = K^{-1} \boldsymbol{\mathcal S}({\bfth}_0) + K^{-1} 
\mathcal{\mathbf{H}}(\bfth^{\ast}) ({\hat{\bfth}}-{{\bfth}_0}),
$
which implies
\begin{eqnarray}
({\hat{\bfth}}-{{\bfth}_0}) &=& \l[ K^{-1} \mathcal{\mathbf{H}}(\bfth^{\ast})      \r]^{-1}  K^{-1} [ \boldsymbol{\mathcal S}
(\hat{\bfth})-\boldsymbol{\mathcal S}({\bfth}_0)] \nn \\
 &=& O_P(1) \l\{- K^{-1} \boldsymbol{\mathcal S}({\bfth}_0)  +    O_P\l( m_K^{-1}  \r)      \r\}, \label{Eq. hat_zero}
\end{eqnarray}
where, in the second line, we use the LLN which implies $\l[ K^{-1} \mathcal{\mathbf{H}}(\bfth^{\ast})  \r]^{-1} K^{-1} = O_P(1)$, and make use of 
(\ref{Eq. Shat}). Then, by standard CLT, we have 
$K^{-1} \boldsymbol{\mathcal S}({\bfth}_0)=O_P\l(K^{-1/2} \r)$. Thus, 
\begin{eqnarray}
({\hat{\bfth}}-{{\bfth}_0}) &=& O_P(1) \l[O_P (K^{-1/2}) +    O_P\l( m_K^{-1}  \r)      \r] \nonumber  \\
&=& O_P\l( g_K \r), \quad \text{with} \quad g_K=\max\l(K^{-1/2};m_K^{-1}\r). \label{Eq. Op}
\end{eqnarray}
%

Thus, the GLAMLE has a rate $g_K$ which depends on both $K$ and $m$, as in (\ref{Eq. Op}). Intuitively, the root-$K$
term comes from standard asymptotic theory of M-estimators, whereas the $m^{-1}$ term comes from the
Laplace approximation.  Looking at (\ref{Eq. Op}), we notice that, if the number of network views ($K$) and the number of edges ($m$) are such that $m_K = O(K^{\varrho_1})$, with {$0<\varrho_1 \leq 1/2$},
then the GLAMLE has rate $O_P(m_K^{-1})$. This seems particularly suitable for the statistical analysis of big datasets containing many views ($K$ is large) of a large graph (namely, a graph having a large number of edges $m$).  If $\varrho_1 > 1/2 $, the Laplace approximation to the actual density is such that the GLAMLE has the same (root-$K$) rate of convergence to ${\bfth}_0$ as the QMLE to the pseudo-true value; see section \ref{Sec: Asy1}.

\section{Monte Carlo exercises} \label{Sec: MC_gllvm}

\subsection{Design and computational aspects}

To illustrate numerically the performance of the GLAMLE, we consider several Monte Carlo experiments. 
We focus on Bernoulli random variables for a simulated directed network with $n_V=18$ and we study the estimation problem of an adjacency matrix, as in Section \ref{Sec: adj}. We consider 1000 Monte Carlo runs and we study the behavior of the GLAMLE under the dyad dependence Assumption \textbf{A2} (unit variance) and \textbf{A2$'$} (non unit variance). In each Monte Carlo run, we set  $K = 100$, namely, we have 100 snapshots of the network. To investigate on the role of the number of latent variables, we consider the cases of one ($q=1$) and two ($q=2$) latent variable/s $\mathbf{Z}$,  under both \textbf{A2} and \textbf{A2$'$}. The goal of the numerical exercises is to study the bias due to the likelihood approximation and the variability of the estimates when the covariance matrix of the latent variables is not the identity matrix---the diagonal elements are greater than one.

Some computational details for an efficient calculation of the GLAMLE are in order. To implement our Laplace approximation method we proceed similarly to \citet{Nikuetal2019}. Specifically, we suitably adapt to our graph setting the \texttt{R} package \texttt{gllvm} (see \citet{N17}) developed for the analysis of multivariate data. The package makes use of
 automated differentiation,  as implemented in the TMB software of \citet{Kr15TMB},  which offers a general tool for the statistical analysis of complex  models, and it is based on \texttt{C++}. Such general software is extremely convenient to calculate  the  Laplace approximation in (\ref{Eq: ellLap}), which is a key step for estimation and inference in our GGLLVMs---indeed, the \texttt{gllvm} has stable routines to compute the  Laplace-approximated likelihood in the case of binary responses for multivariate random variables: this code provides us with the stepping stone for the implementation of our GLAMLE in our network setting. We refer to \citet{Nikuetal2019}, p.6 section 3.1, for numerical details; additional aspects are available directly from the \texttt{gllvm} manual. In our estimation method, the Laplace-approximate log-likelihood cannot be easily directly maximized, for example by means of some gradient-based optimization algorithms already available in \texttt{R}, like \texttt{optim}. Indeed, this entails the need for calculating the gradient functions for each dyadic response $Y_{ij}$ separately, leading to a long, computationally demanding and unstable optimization procedure. To overcome this issue, we make use of TMB. Following \citet{Nikuetal2019},  we   write in  \texttt{C++} the Laplace-approximated log-likelihood (templates are available). Then, we exploit the fact that TMB employs the \texttt{C++} library \texttt{CppAD} to calculate the gradient and the Hessian. Finally, we call these objects (in fact they are outputted as functions) from \texttt{R} and we make use of them into gradient based optimization method---see  \texttt{gllvm} for computational details about the starting values, which we control via the default \texttt{``res''} option. In our experience, this procedures yields a stable, efficient and fast routine.

\subsection{Estimation results: bias and variance}  \label{Sec: A2A21}
\begin{figure}[htp!]
\begin{center}
\begin{tabular}{cc}  \hspace{0.5cm}$q=1$& \hspace{0.5cm} $q=2$ \\
\includegraphics[width=0.4\textwidth, height=0.23\textheight]{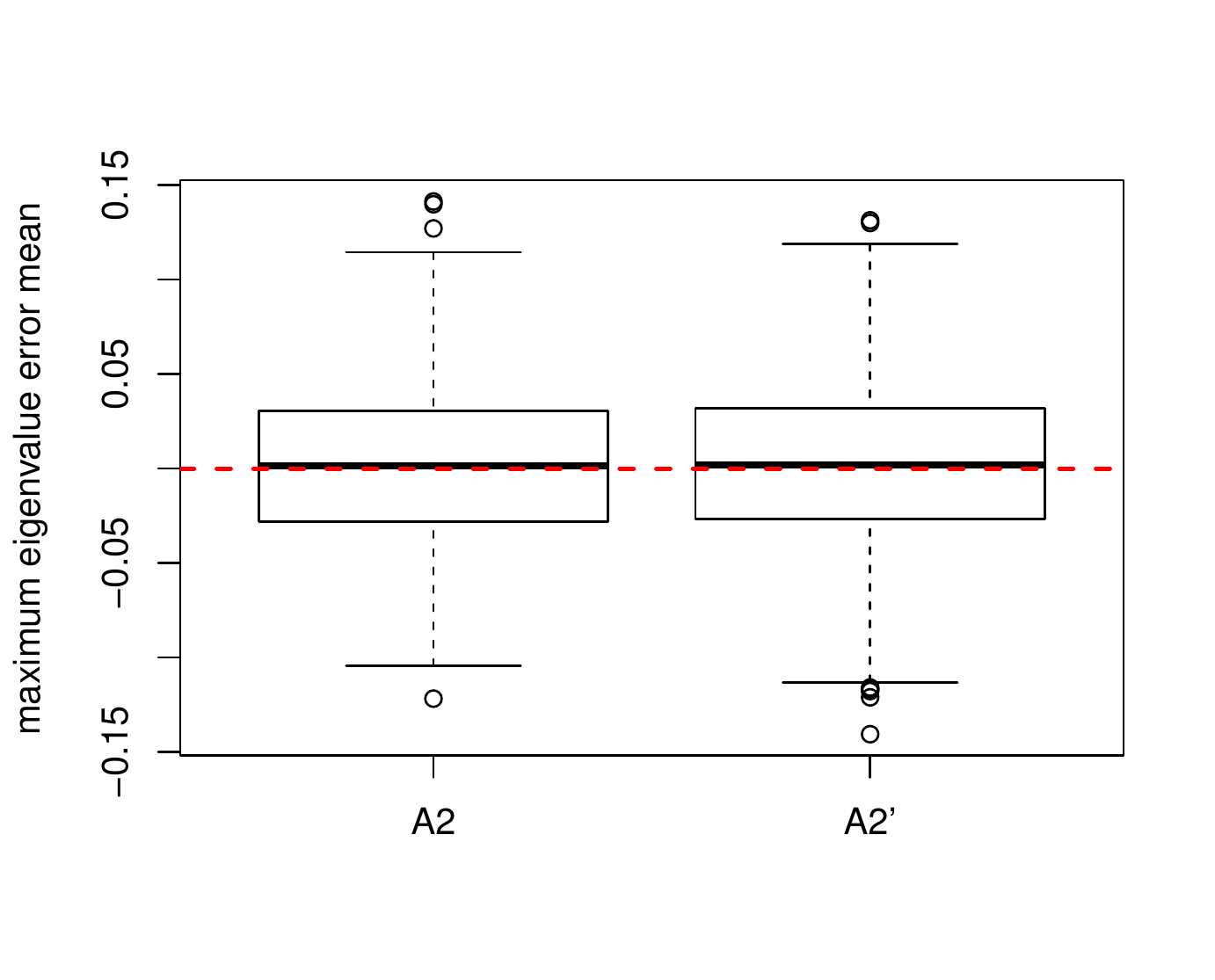}&
\includegraphics[width=0.4\textwidth, height=0.23\textheight]{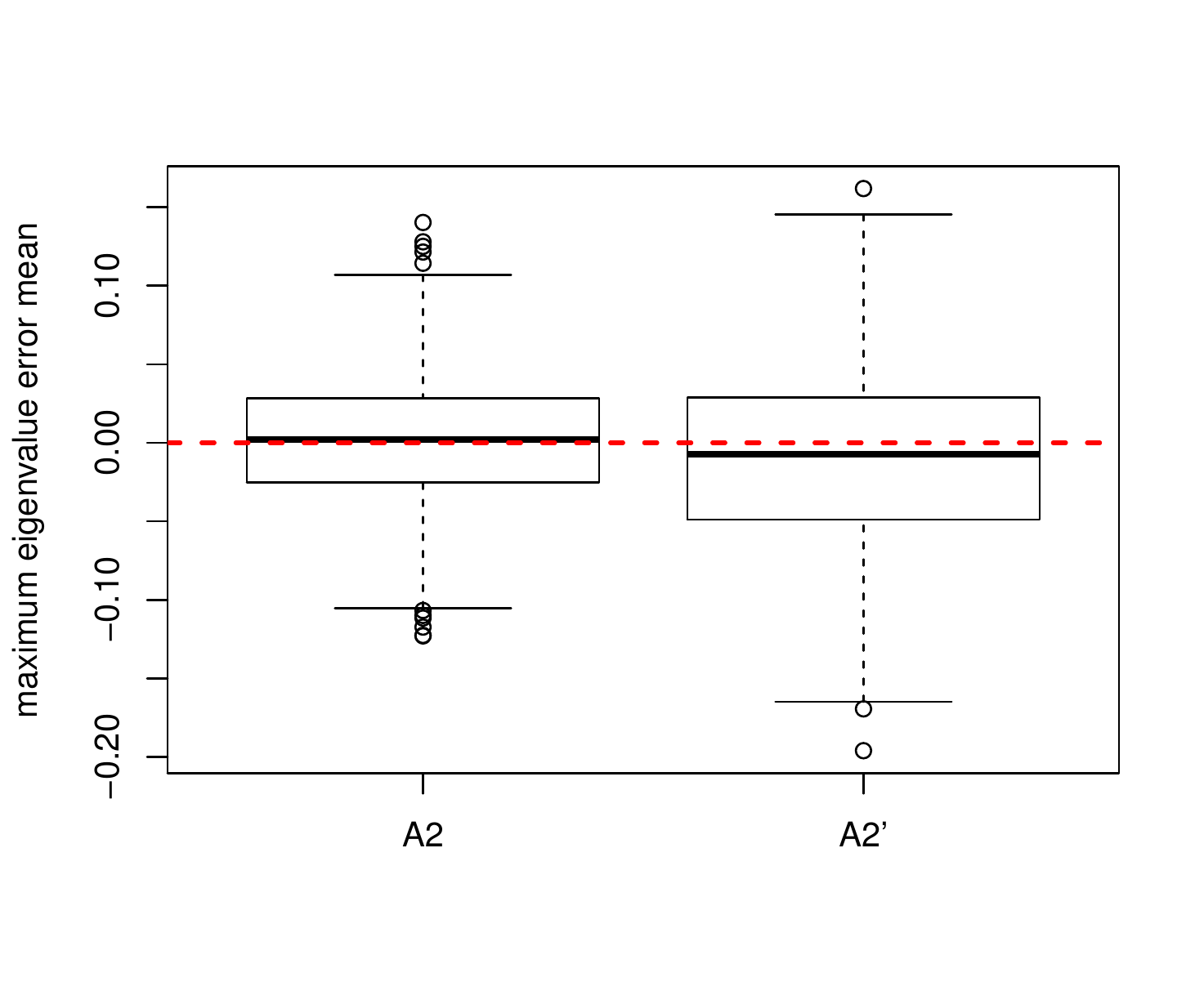}
\end{tabular}
\caption{ GLAMLE under  \textbf{A2} and \textbf{A2$'$}. 
 Boxplot of error means of maximum eigenvalues of  $\hat{\boldsymbol{\pi}}$ and $\boldsymbol{{\pi}}_{0}$ for $q=1$ (left panel) and $q=2$ (right panel). Directed graph with 18 nodes, $K=100$ sample size, Monte Carlo size 1000.}
    \label{Fig: Mean}
\end{center}
\end{figure}

To investigate the performance of our GLAMLE, 
we compare  $\hat{\boldsymbol{{\pi}}}$ with the true $\boldsymbol{{\pi}}_{0}$, representing the estimated and the true probability for each element of the adjacency matrix, respectively. More precisely, each $\pi_{ij}$ is given in (\ref{Eq: Pyij}) and $\hat\pi_{ij}$ is the corresponding estimator obtained with a plug in of the GLAMLE estimated parameters. Since we work on the estimation of the entries of a $n_V\times n_V$-matrix, we consider several metrics to assess the performance of the GLAMLE, under the different estimation settings. First, as it is common in multivariate analysis, we focus on the  maximum eigenvalue of the matrix. Specifically, we work on the error means of the true and estimated  maximum eigenvalue, which is defined as follows. For each $k$-th layer, let $\hat{ei}^{(k)}$ and ${ei_0}^{(k)}$ represent the maximum eigenvalue of $\hat{\boldsymbol{{\pi}}}^{(k)}$ and of $\boldsymbol{{\pi}}_0^{(k)}$, respectively. Then, the error is $\hat{ei}^{(k)}-{ei_0}^{(k)}$ and the error mean is
${K}^{-1}\sum_{k=1}^{K}\l(\hat{ei}^{(k)}-{ei_0}^{(k)}\r).$ 
This measure 
gives information about the mean of the estimation error (which is related to the estimation bias) and of its variability:  a comparison of the defined error mean under \textbf{A2} and \textbf{A2$'$} is informative about the average performance of the GLAMLE. In Figure \ref{Fig: Mean}, we display the related results. We see that the boxplots are centered around zero: this illustrates that the GLAMLE gives on average good estimates, displaying a small bias. Elaborating further, let us consider the case of one latent variable ($q=1$).  The left panel  illustrates that the boxplots are similar under \textbf{A2} and \textbf{A2$'$}: the only difference between the two settings is that the variance of $\mathbf{z}_{(2)}^{(k)}$ is unit (under \textbf{A2})  or not (under \textbf{A2$'$}). The plots provides evidence that the GLAMLE performs essentially in the same way under the two settings. When $q=2$, the right panel of Figure \ref{Fig: Mean} shows that the GLAMLE displays a small bias under \textbf{A2$'$}, whilst it remains unbiased under \textbf{A2}. This is certainly due to the number of views ($K=100$) which is a moderate to small size: additional, unreported, Monte Carlo experiments illustrates that the bias becomes negligible for larger values of $K$.

\begin{figure}[htp!]
\begin{center}
\begin{tabular}{cc} \hspace{0.5cm}$q=1$& \hspace{0.5cm} $q=2$ \\
\includegraphics[width=0.4\textwidth, height=0.18\textheight]{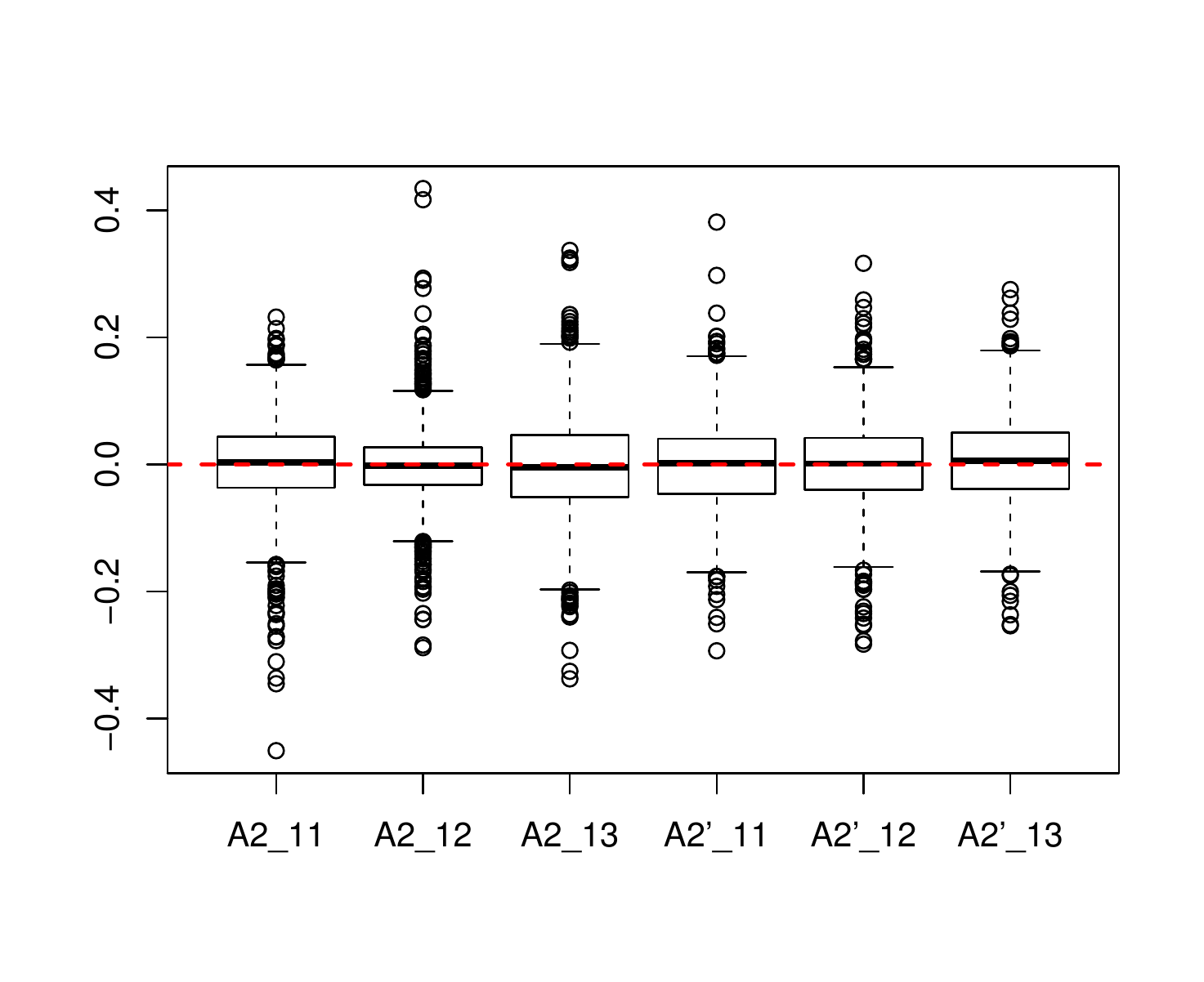}&
\includegraphics[width=0.4\textwidth, height=0.18\textheight]{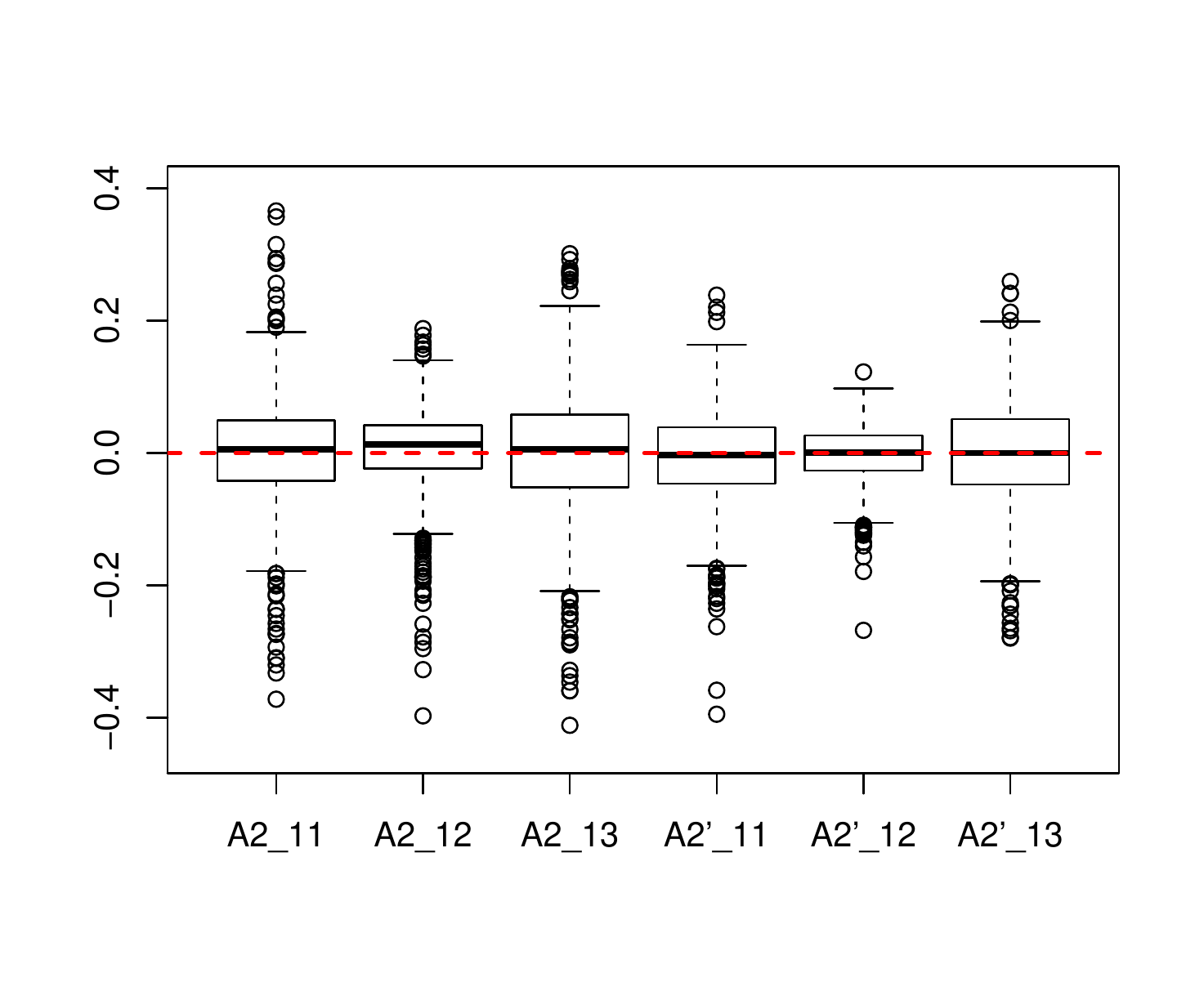}\\
%
\includegraphics[width=0.4\textwidth, height=0.18\textheight]{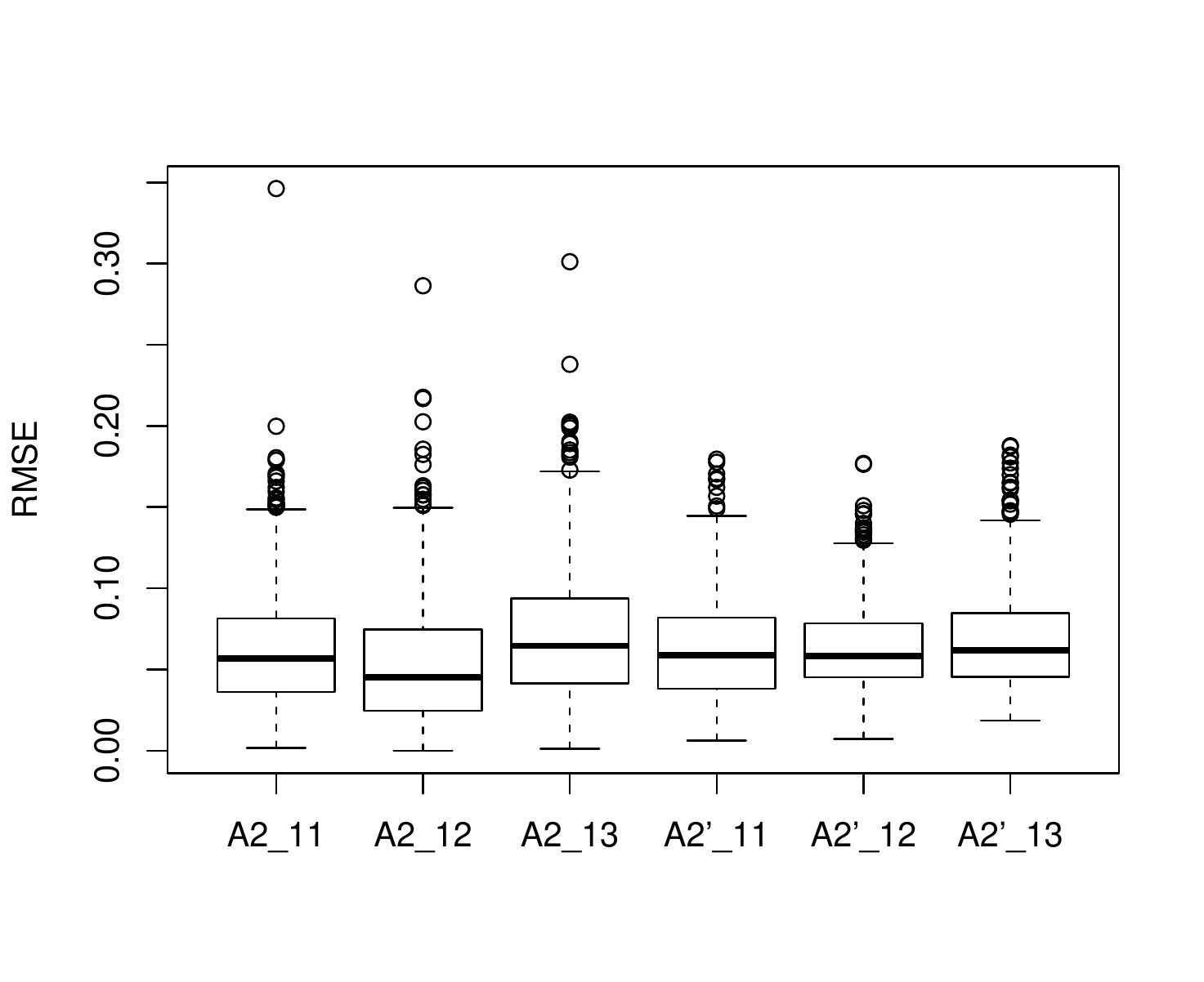}&
\includegraphics[width=0.4\textwidth, height=0.18\textheight]{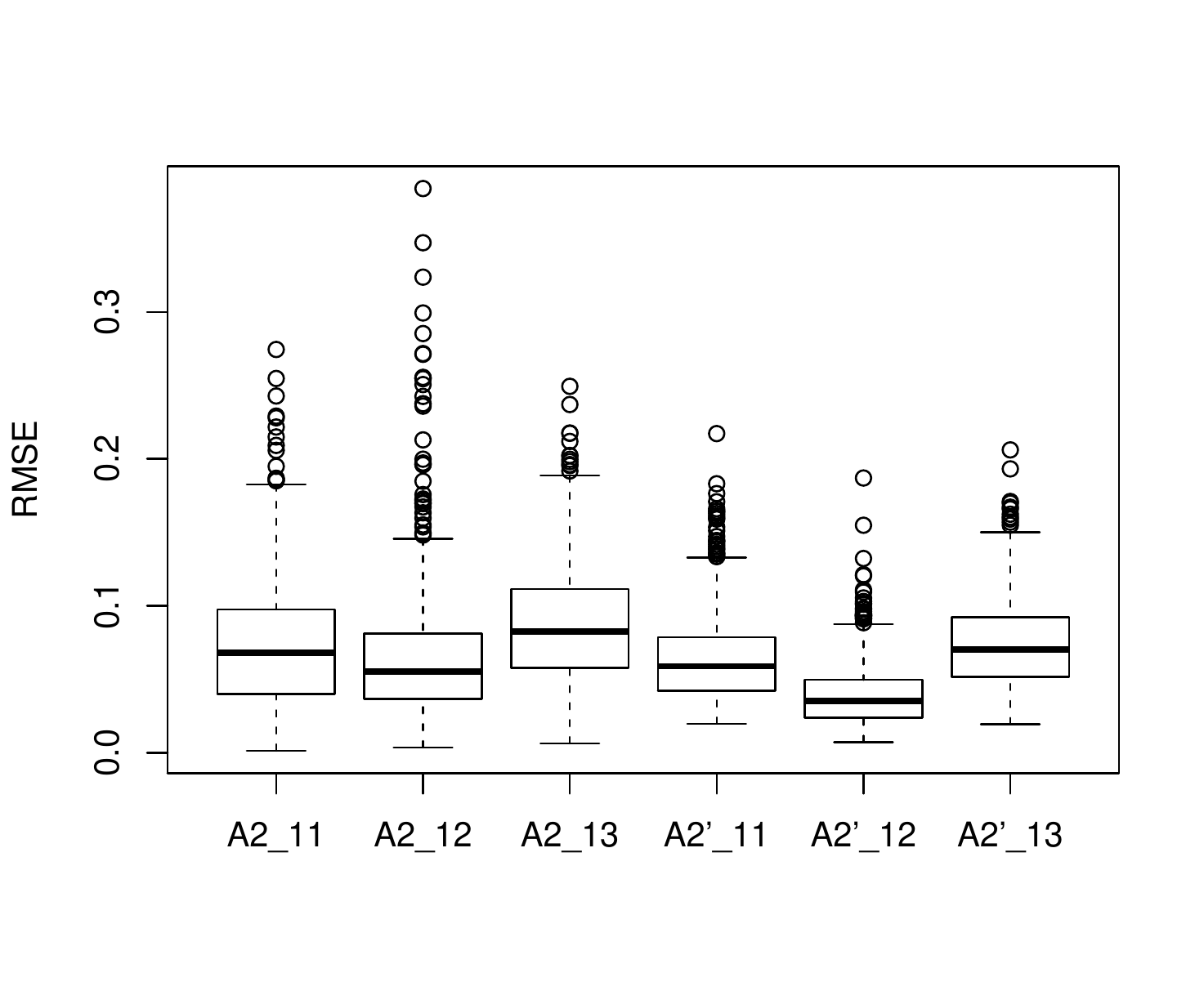}\\
\end{tabular}
\caption{GLAMLE under \textbf{A2} and \textbf{A2$'$}, for one ($q=1$, left panels) and two ($q=2$, right panels) factors. Top panels: 
Boxplots of $\hat{\pi}^{(1)}_{ij}-\pi^{(1)}_{ij}$ for three different edges $ij = \{11,12,13\}$.  Bottom panels: Boxplots of RMSEs of  $\hat{\pi}_{ij}$ for the same edges.  Directed graph with 18 nodes and $K=100$, Monte Carlo size 1000.}
   \label{Fig: RMSE}
\end{center}
\end{figure}

To investigate further, 
we look at the element-wise differences and we introduce 
the RMSEs of $\hat\pi_{ij}$, across $K$ observations, computed as 
$$
\text{RMSE}(\hat\pi_{ij}) = \sqrt{{\sum_{k=1}^{K}\l(\hat\pi_{ij}^{(k)}-\pi_{ij}^{(k)}\r)^2}{K}^{-1}}, 
$$
For the sake of visualization, in Figure \ref{Fig: RMSE} we display the boxplots for $(\hat\pi^{(1)}_{ij}-\pi^{(1)}_{0,ij})$ for three edges $ij = \{11,12,13\}$---very similar results are available for other edges, unreported. For the same set of edges, we display the RMSEs: the boxplots illustrate that the GLAMLE, under $q=1$ and $q=2$, has similar performance under  \textbf{A2} and \textbf{A2$'$}. 

\color{black}

\subsection{Comparisons with other methods} \label{Sec: VA}

Markov chain Monte Carlo (MCMC) and Variational approximation (VA) approaches are widely used for estimating LVMs. 
In this section, we compare the performances between Laplace and other estimation methods (MCMC and VA) for a Bernoulli GGLLVM with 18 nodes under Assumption \textbf{A2}. We remark that, since our modeling framework is original, a viable MCMC procedure is not currently available from the literature. Thus, we implemented one following other similar approaches available from the latent position models literature \citep{HRH02,rastelli2018computationally}; implementation details are provided in Appendix \ref{app:mcmc}, and the code is available upon request. Similarly, there is no VA approach already available for our model: in our implementation, we adapted the procedure introduced by \citet{HWOH17}, which is implemented in \texttt{gllvm}---see the package manual.

\begin{figure}[htbp!]
\begin{center}
\begin{tabular}{cc}  \hspace{0.5cm}$q=1$& \hspace{0.5cm} $q=2$ \\
\includegraphics[width=0.4\textwidth, height=0.23\textheight]{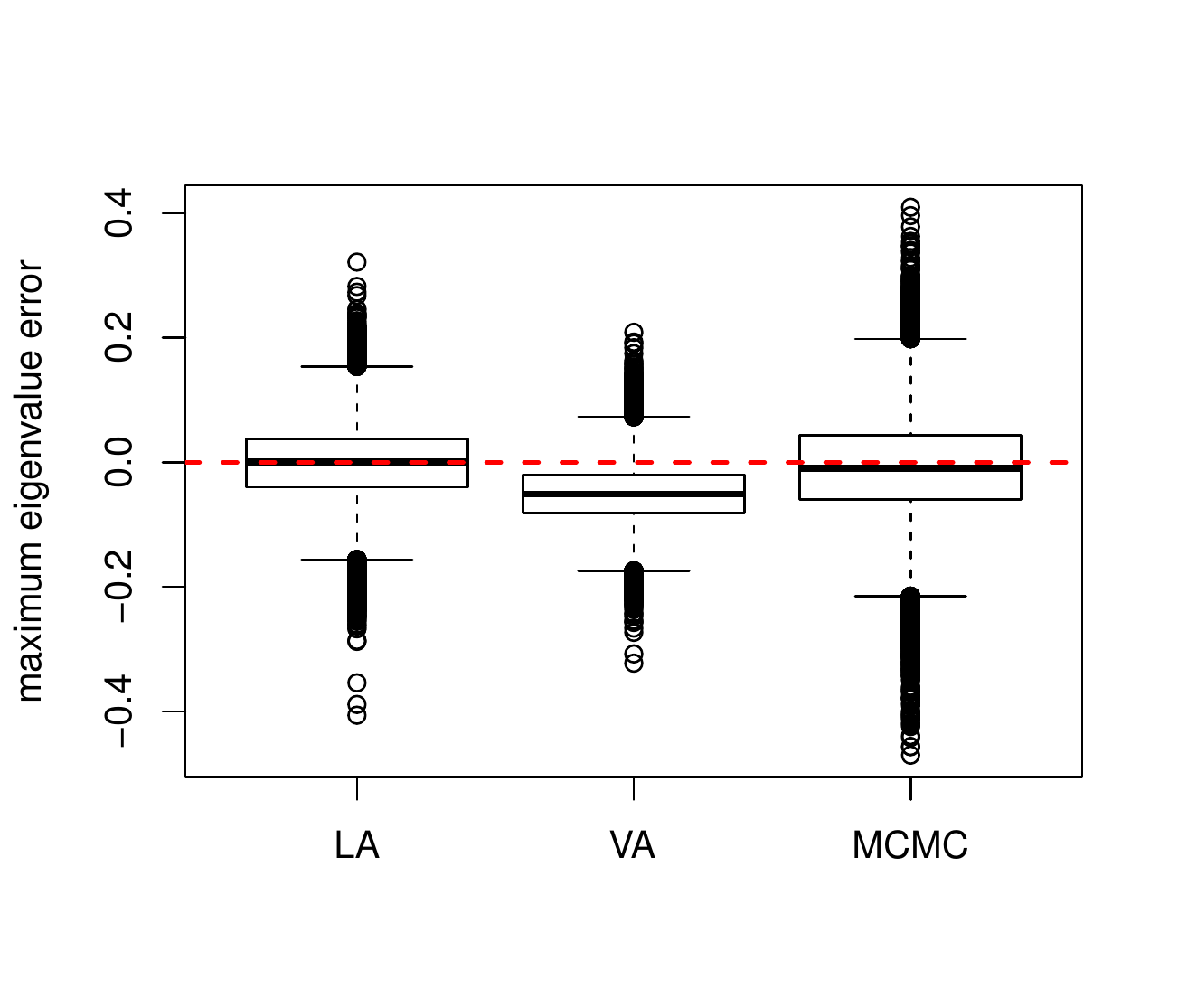}&
\includegraphics[width=0.4\textwidth, height=0.23\textheight]{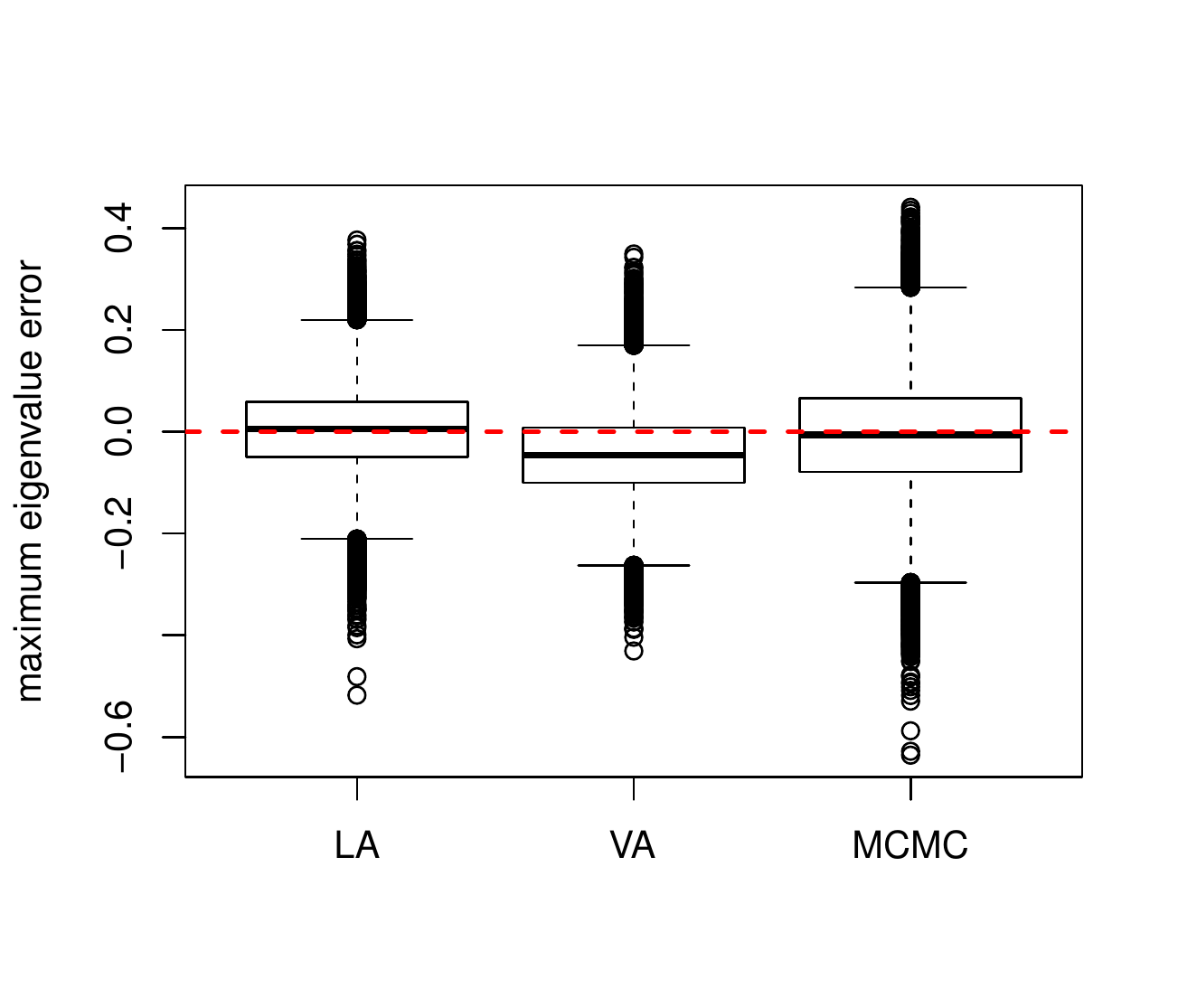}
\end{tabular}
\caption{ Comparison between Laplace approximations (LA), variational approximations (VA) and Markov chain Monte Carlo (MCMC) on a Bernoulli GGLLVM under Assumption \textbf{A2}. 
Left Panel: Boxplot of error of maximum eigenvalues of  $\hat{\boldsymbol{\pi}}$ and $\boldsymbol{{\pi}}_{0}$ using one latent variable. Right Panel: Boxplot of error of maximum eigenvalues of  $\hat{\boldsymbol{\pi}}$ and $\boldsymbol{{\pi}}_{0}$ using two latent variables. We consider a directed graph with 18 nodes and 100 sample size. Monte Carlo size is 1000.}
    \label{Fig: Mean_VA}
\end{center}
\end{figure}

To evaluate the performance of the  estimation methods, first, we plot the differences between the true and estimated maximum eigenvalues of $\hat{\boldsymbol{{\pi}}}$ and $\boldsymbol{{\pi}}_{0}$. Figure \ref{Fig: Mean_VA} displays the results. The plot confirms that the LA has small bias with values centering around $0$. Large biases appear with VA method: this is expressed by the fact that the median for VA is below $0$.  Moreover,  we see that MCMC provides accurate estimations, but it has a slightly larger interquartile range than LA and more outliers than the other methods.

To gain further insights into the behavior of LA, VA and MCMC, we consider an undirected graph with 18 nodes at time point $k$, and compare its true and estimated conditional probabilities $\boldsymbol{\pi}^{(k)}$ for each edge. 
\begin{figure}[htbp!]
\begin{center}
\begin{tabular}{cc}  \hspace{0.5cm}$q=1$& \hspace{0.5cm} $q=2$ \\
\begin{turn}
{90} \hspace{3cm} LA 
\end{turn}
\includegraphics[width=0.4\textwidth, height=0.24\textheight]{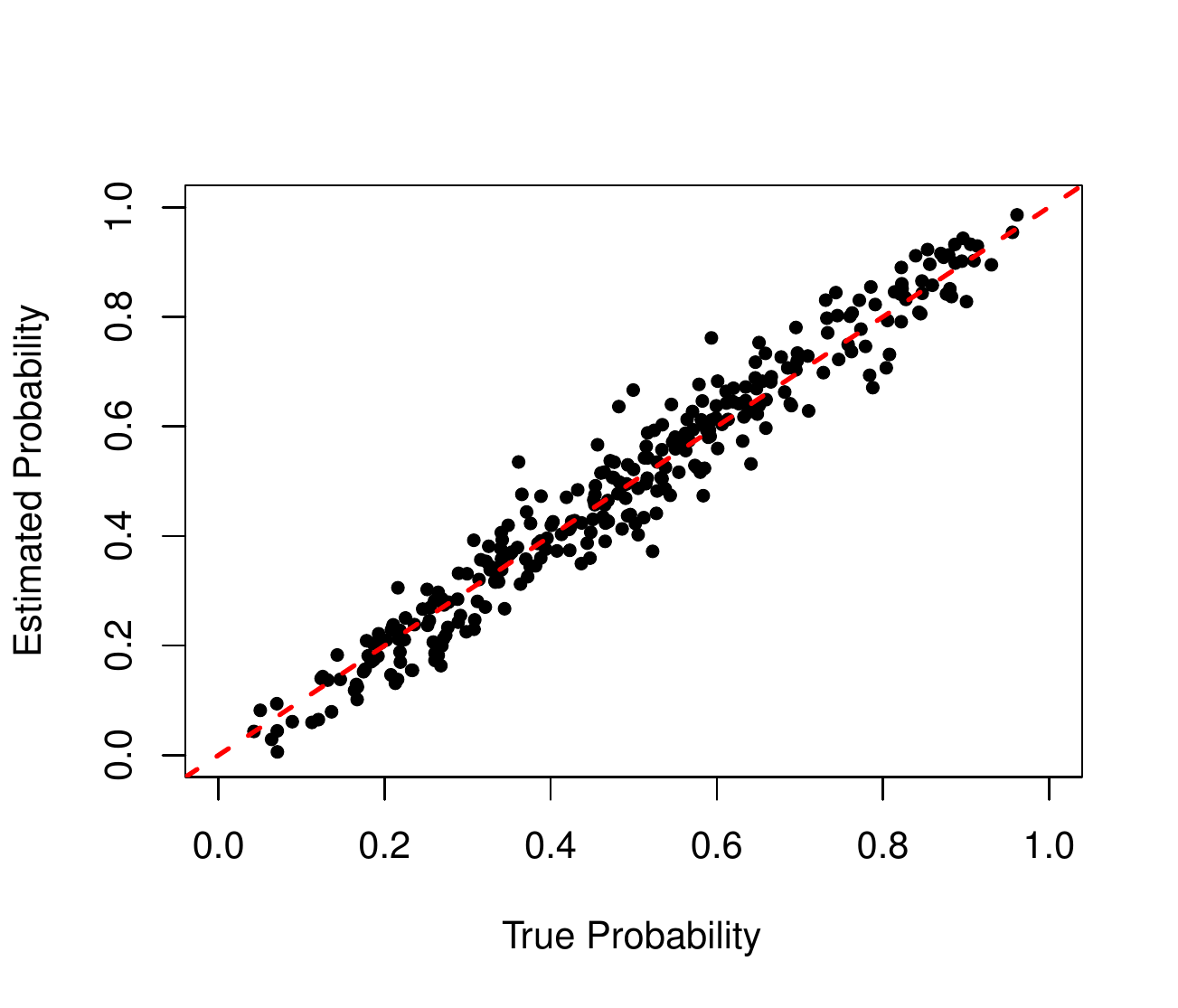}&
\includegraphics[width=0.4\textwidth, height=0.24\textheight]{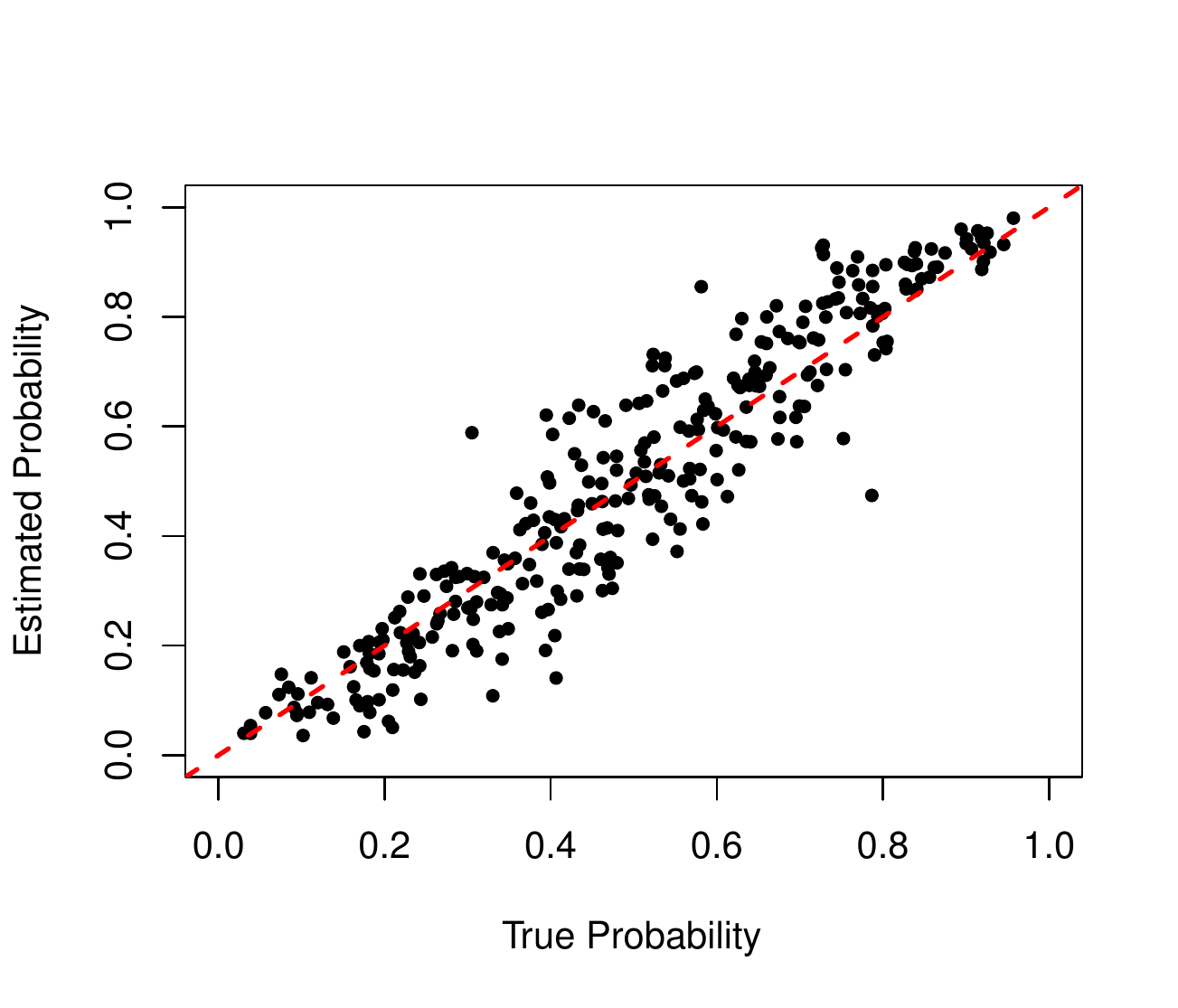}\\
\begin{turn}
{90} \hspace{3cm} VA
\end{turn}
\includegraphics[width=0.4\textwidth, height=0.24\textheight]{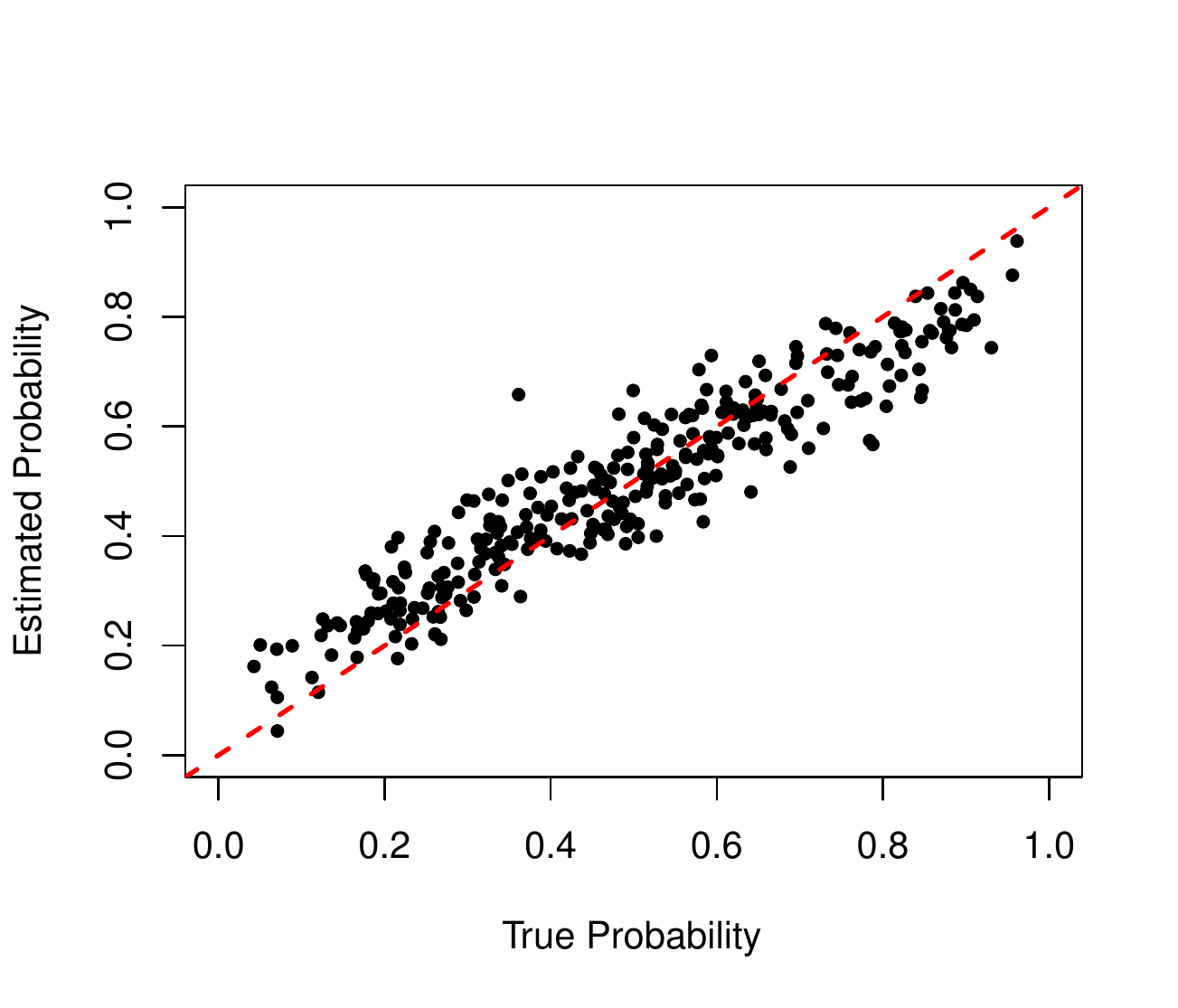}&
\includegraphics[width=0.4\textwidth, height=0.24\textheight]{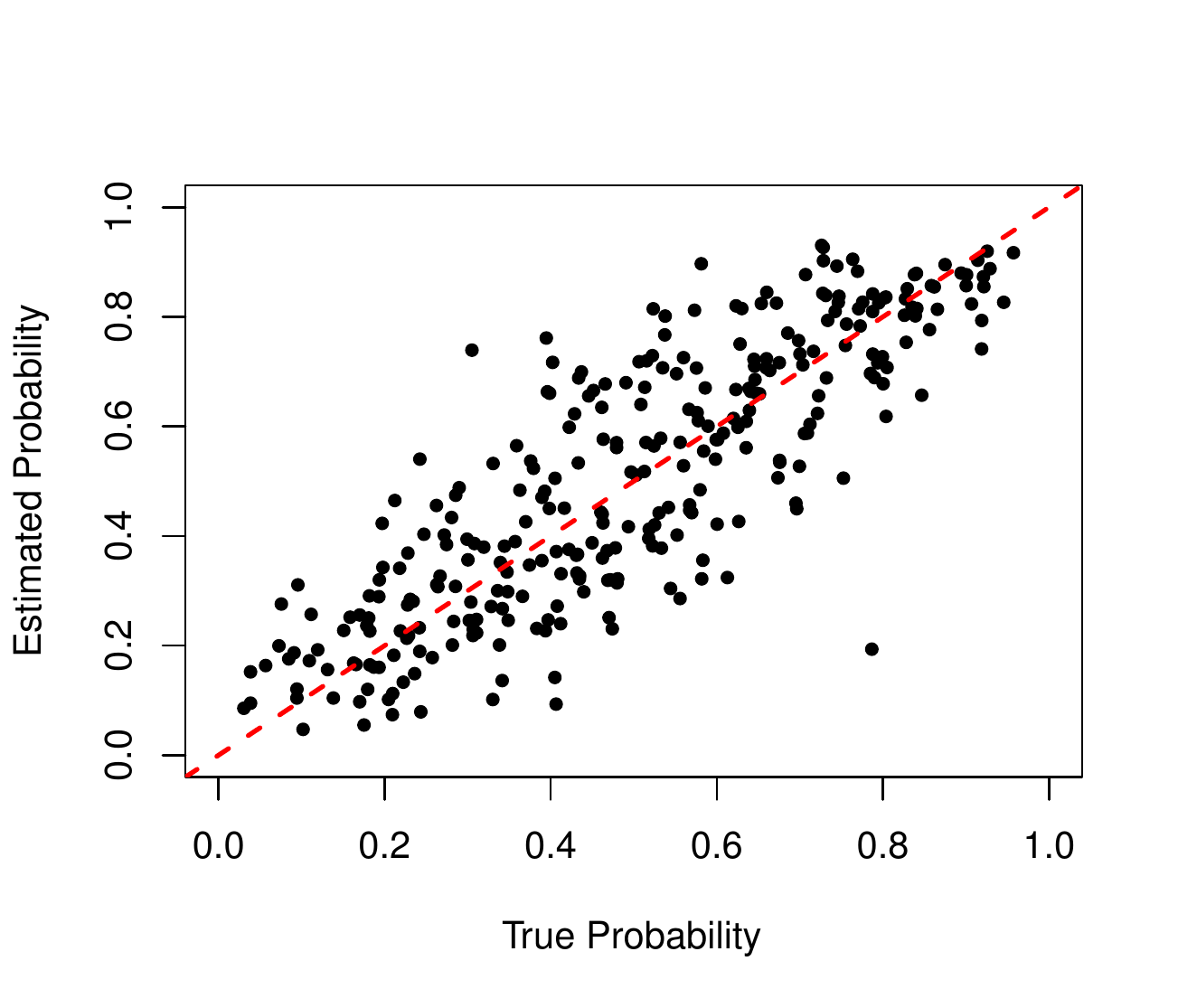}\\
\begin{turn}
{90} \hspace{3cm} MCMC
\end{turn}
\includegraphics[width=0.4\textwidth, height=0.24\textheight]{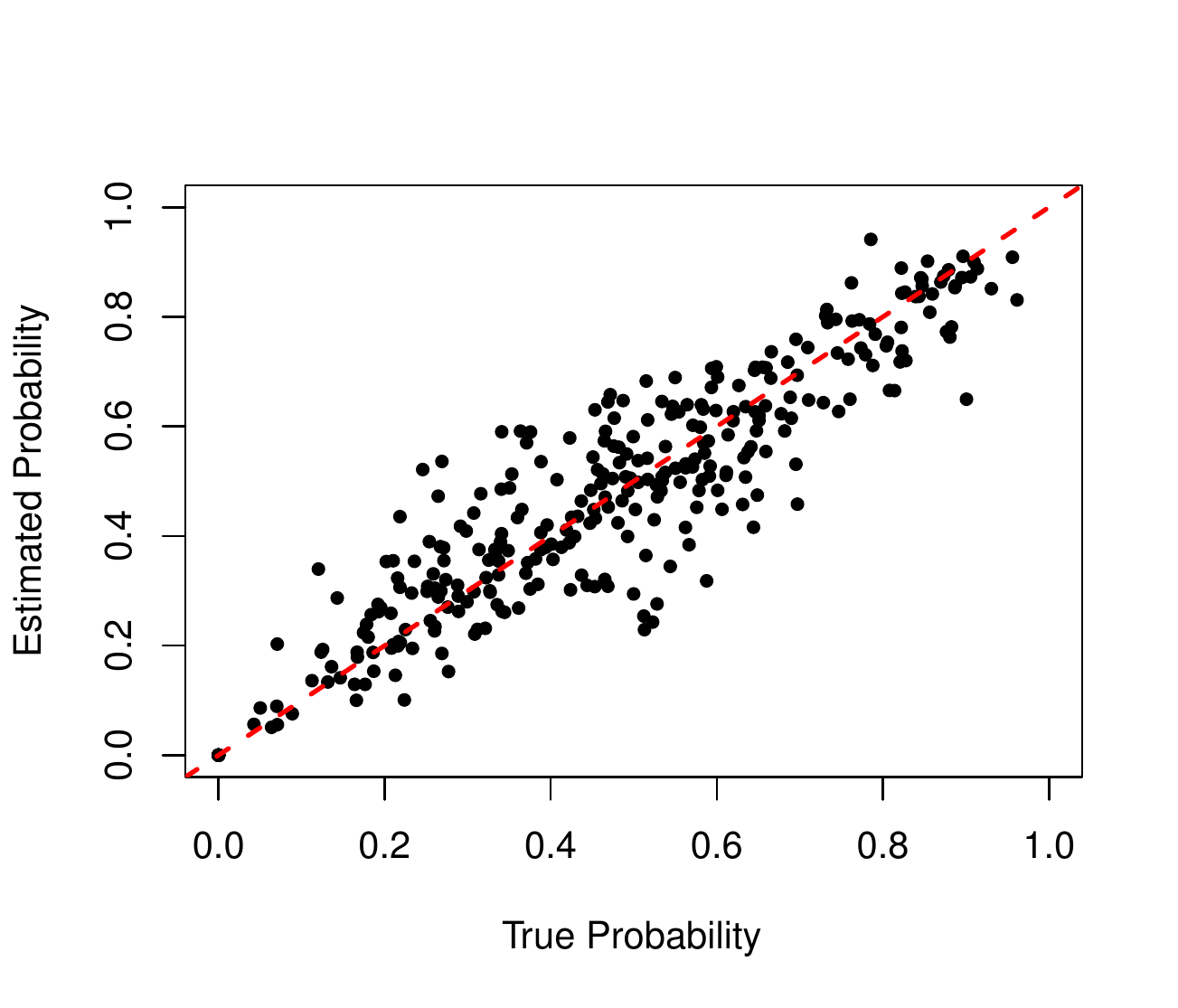}&
\includegraphics[width=0.4\textwidth, height=0.24\textheight]{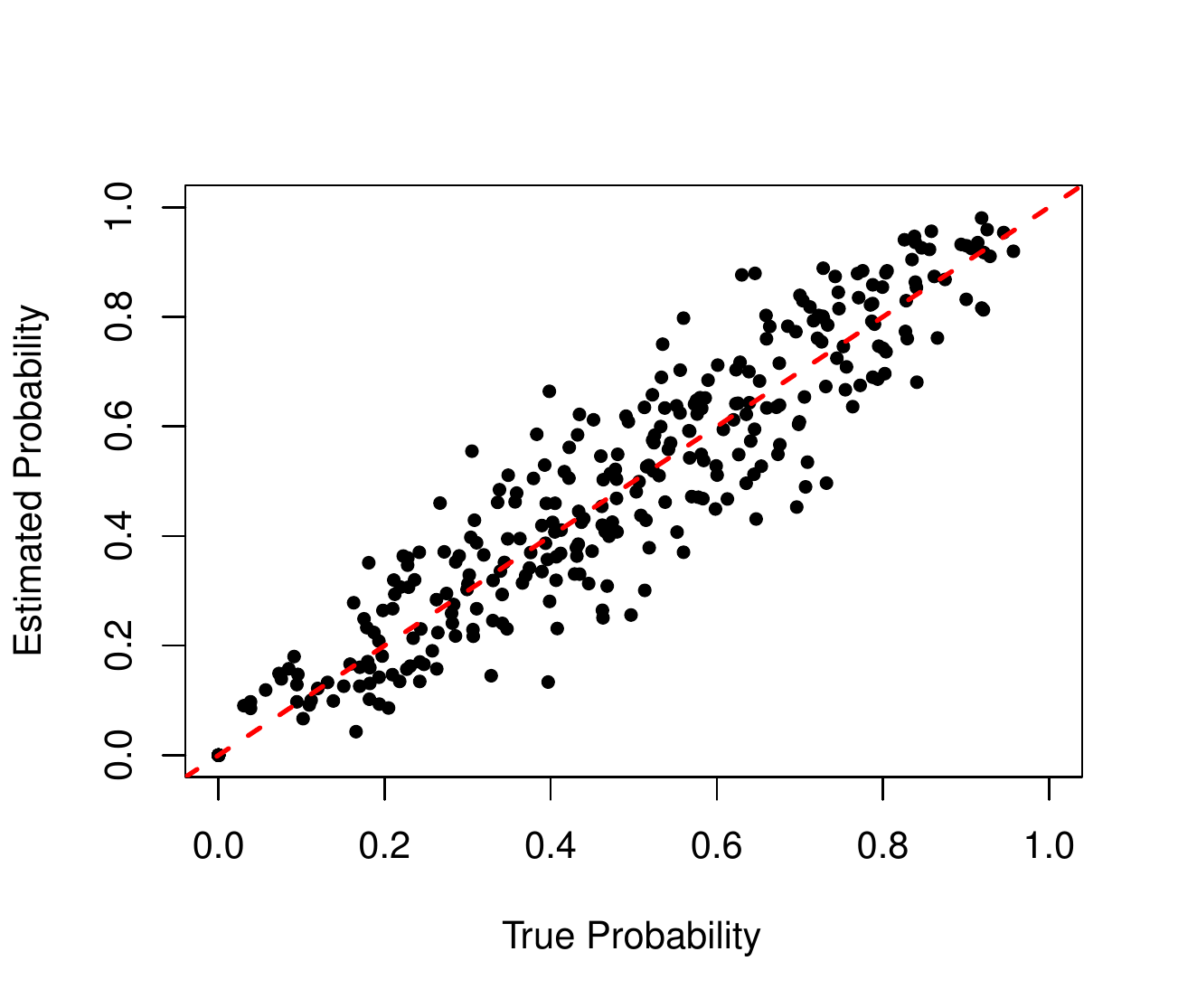}
\end{tabular}
\caption{Bernoulli GGLLVM with one or two latent variable using Laplace approximation (LA), variational approximation (VA) and Markov chain Monte Carlo (MCMC), under Assumption \textbf{A2}. Comparison between true and estimated $\boldsymbol{\pi}^{(1)}$. We consider a directed graph with 18 nodes and 100 sample size.}
    \label{Fig: pi_VA}
\end{center}
\end{figure}
In Figure \ref{Fig: pi_VA}, we display the related results. Also in this case, the LA outperforms VA and MCMC,  as it has a smaller variation compared to the other two, and estimates align better with the 45 degrees line.  For $q=1$,  we find that VA has a large departure when the probabilities fall over the interval $[0,0.3]\cup [0.7,1]$, corresponding to an overestimation in the lower tail, and to an underestimation  in the upper tail, respectively.  In contrast, LA and MCMC provide accurate estimates (the dots are close to the 45 degrees line,  also in the tails).
Finally, we study the computation times for the three methods. In all experiments, we see that VA is faster than the other two methods, MCMC is the slowest method and LA is in between the other two, but it remains much faster than MCMC.
For instance, when $q = 1$,  the computation time of MCMC (on a 3 GHz Intel Core i7
processor) is about 6 minutes,  two times slower than LA (computation time about 2.5 minutes), and much slower than VA (computation time about 15 seconds). In all experiments, the gain in speed yielded by VA comes with a cost in terms of accuracy. Thus, we conclude that  LA guarantees an excellent balance between speed and accuracy.

 \color{black}

\section{Real data motivating example (reprise)}\label{Sec: application}

\subsection{Data summary} 

\begin{figure}[hbtp]
\begin{center}
\includegraphics[width=0.4\textwidth, height=0.24\textheight]{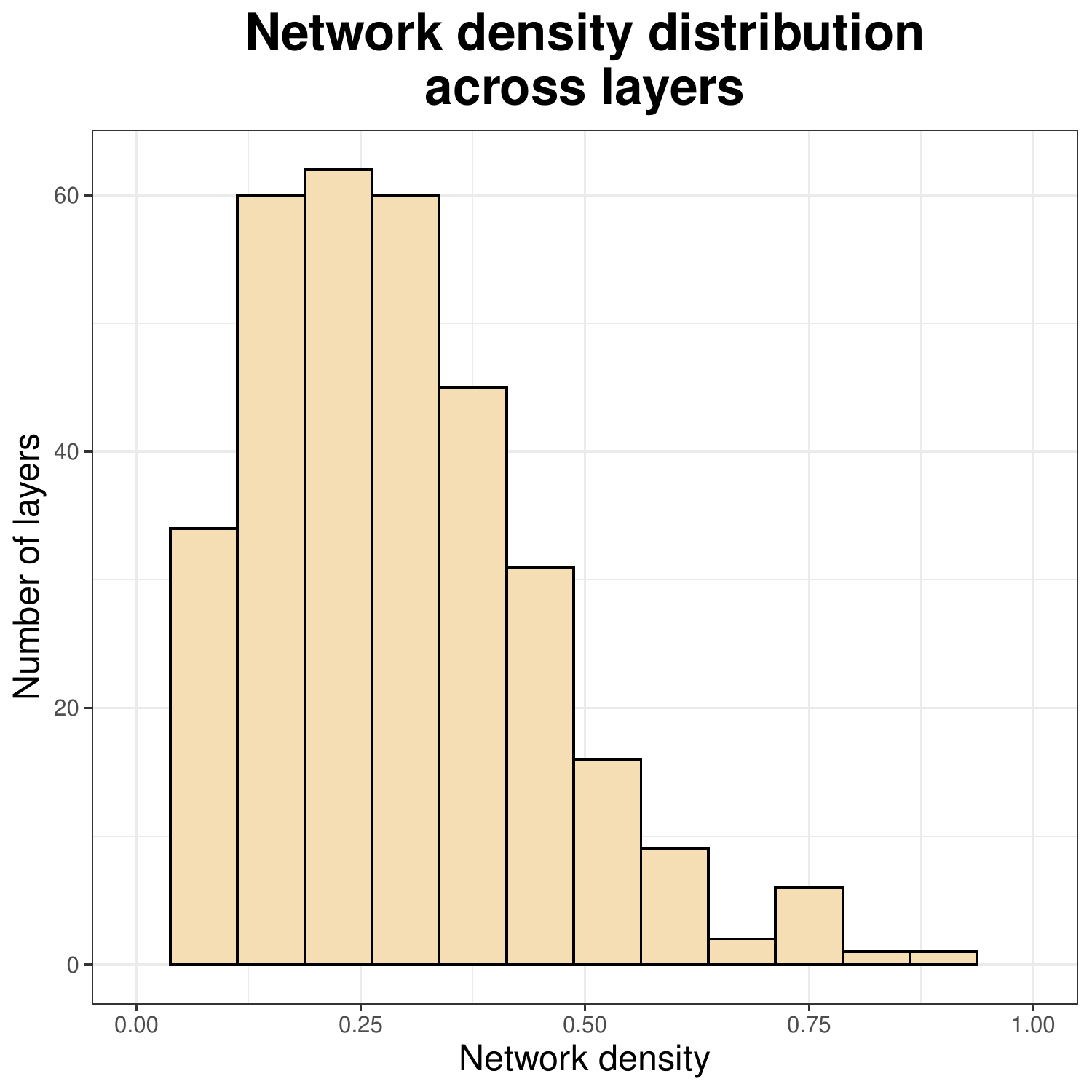}
\includegraphics[width=0.4\textwidth, height=0.24\textheight]{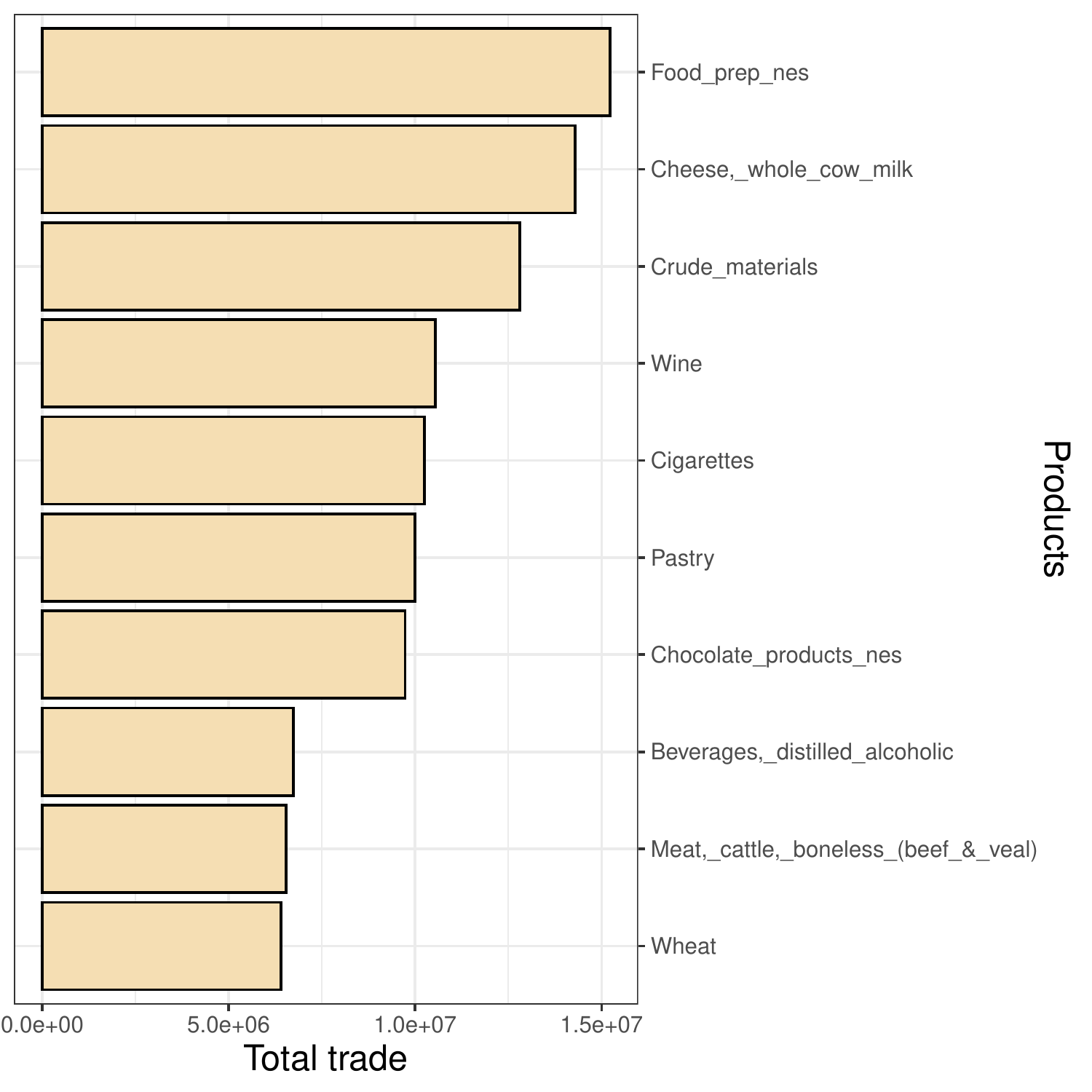}
\caption{FAO network. The left panel shows the network densities (defined as the proportion of edges that appear in a network) for all the products. The right panel shows instead the total traded volume for the most traded 10 products (value in $1{,}000$ US dollars).}
\label{fig:layer_info}
\end{center}
\end{figure}

We consider the multiplex network of the FAO dataset, introduced in Section \ref{realexamples}. For this dataset, we have $K=364$ layers available, each corresponding to a traded product, and we assume that these layers are conditionally independent given the latent variables of our model. We represent each $k$-th layer through a directed binary network, whereby an edge appears from $i$ to $j$ if $i$ has exported product $k$ to $j$. The number of nodes, i.e. countries, in the network is $n_V = 28$ and the number of dyads is $m = 756$.  In Table \ref{tab:number_traded_per_country}, we show the list of countries and the number of different products that each country trades. 
The table highlights that most countries tend to trade most products, meaning that they have at least one trading partner for each traded product.
By contrast, if we look individually at each layer, we see that the individual networks tend to be rather sparse. This makes sense since we expect countries to get access to most products, either through import, or through its own production (thus the country would likely export that product).  This is illustrated in the left panel of Figure \ref{fig:layer_info}, where we observe that the network density (i.e. the proportion of edges that appear) can be very low for some of the layers. The right panel of Figure \ref{fig:layer_info} shows instead the top 10 traded products, by total traded volume (value in $1{,}000$ US dollars). 

\begin{table}[h!]
\setlength{\tabcolsep}{10pt}
\centering
\begin{tabular}{lrlrlrlr}
\toprule
        Belgium    &     343  &  Austria        &    327  &  Sweden    &     299  &  Luxembourg &     265 \\
        Germany    &     343  &  Czech Republic &    326  &  Ireland   &     297  &  Estonia    &     251 \\
    Netherlands    &     341  &  Poland         &    321  &  Slovakia  &     297  &  Slovenia   &     248 \\
         France    &     338  &  Hungary        &    317  &  Lithuania &     294  &  Finland    &     245 \\
          Spain    &     337  &  Denmark        &    313  &  Bulgaria  &     283  &  Croatia    &     175 \\
          Italy    &     336  &  Portugal       &    304  &  Romania   &     276  &  Cyprus     &     173 \\
 United Kingdom    &     335  &  Greece         &    299  &  Latvia    &     275  &  Malta      &      81 \\
\bottomrule
\end{tabular}
\caption{FAO network. Number of different products traded by each of the countries.}
\label{tab:number_traded_per_country}
\end{table}

\subsection{Inference and discussion on the estimation results} 

We fit our GGLLVM under the Assumption \textbf{A2} , with $q=2$. Differently from the simulations, we exclude the intercept term from the model, so that $\textbf{Z}$ and $\boldsymbol{\alpha}_{ij}$ are both two-dimensional vectors. 
We choose this particular setup for visualization purposes, and because it returns a good model fit. However, we emphasize that our choice is arbitrary and more sophisticated procedures could be employed to choose $q$; see Section \ref{Sec: concl}. The output of the procedure consists of the estimates $\hat{\boldsymbol{\alpha}}$ and $\hat{\textbf{z}}^{(k)}$, for $k=1,2...,K$. 
In terms of interpretation, the parameters $\hat{\boldsymbol{\alpha}}$ provide us with model-based general information on the nodes of the network, bypassing the specific features related to each layer. On the other hand, the parameters $\hat{\textbf{z}}^{(k)}$ provide us with model-based information on the different layers of the network, distilling this from the heterogeneous behaviors of the nodes.

\begin{figure}[hbtp]
\begin{center}
\includegraphics[width=0.4\textwidth, height=0.275\textheight]{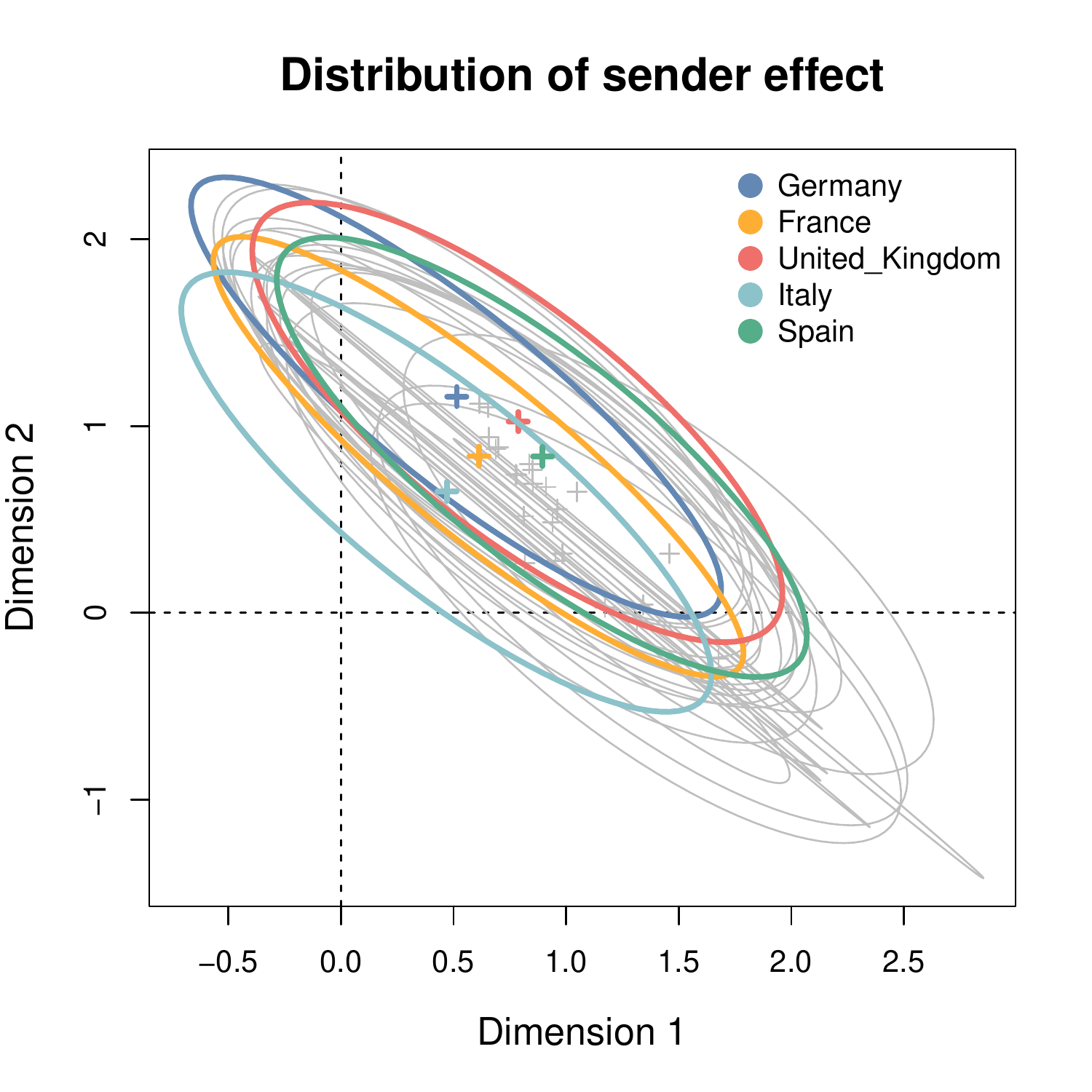}
\includegraphics[width=0.4\textwidth, height=0.275\textheight]{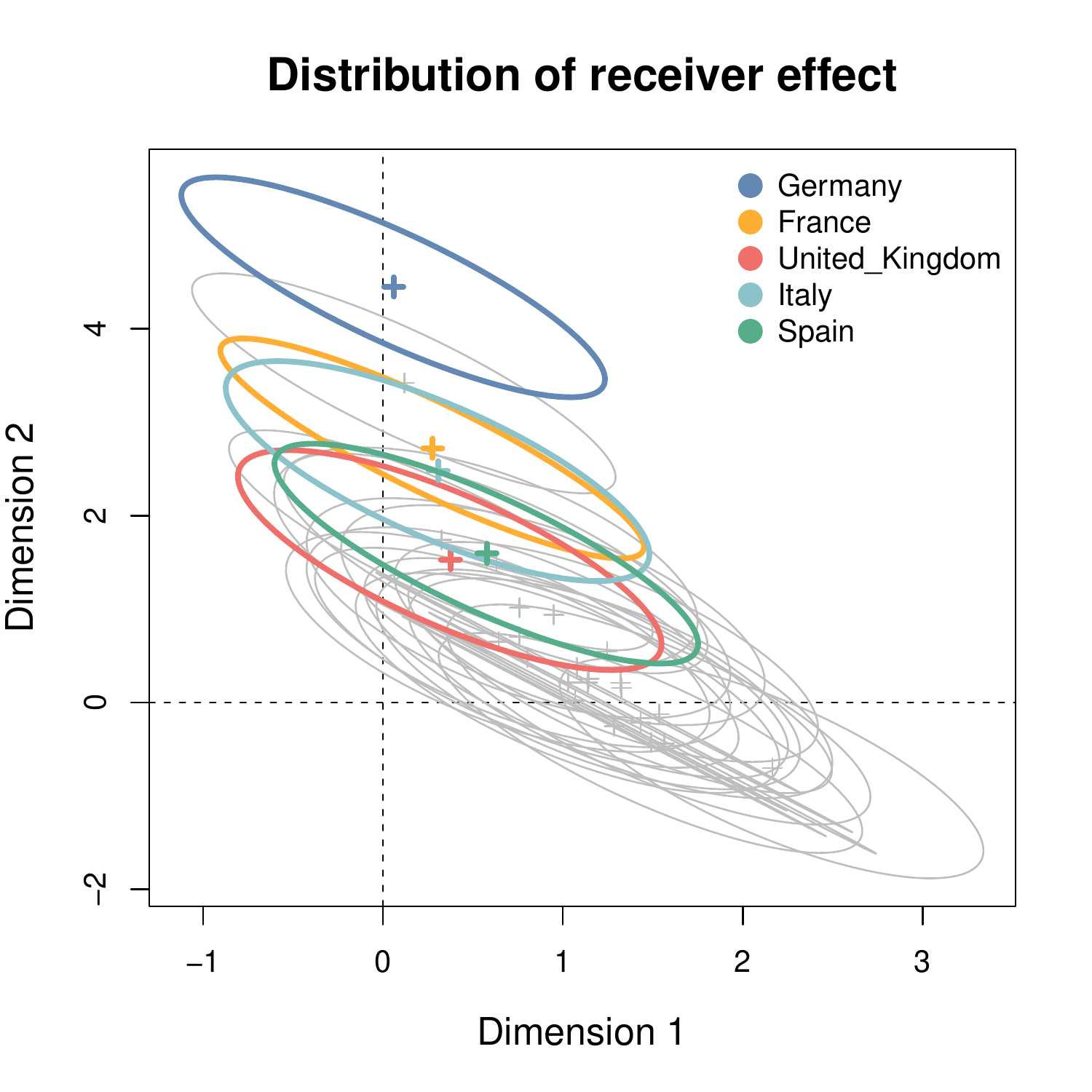}
\caption{FAO network. Sender and receiver effects for the largest EU countries, by population. The ellipses represent the dispersion of the $\boldsymbol{\alpha}_{ij}$ sender values as $j$ varies (resp. receiver values as $i$ varies, on the right panel). The center of each ellipse, represented by a cross, corresponds to the median value.}
\label{Fig:fao_sender_large}
\end{center}
\end{figure}

An arbitrary country $i$ of the network is characterized by two elements: (1) a set of bivariate vectors $\boldsymbol{\alpha}_{i1}, \dots, \boldsymbol{\alpha}_{in_V}$, which determine the tendency of node $i$ to send a connection to any other node in the network; (2) a set of bivariate vectors $\boldsymbol{\alpha}_{1i}, \dots, \boldsymbol{\alpha}_{n_Vi}$, indicating the tendency of $i$ to receive a connection from any other node in the network. 
We note that the $\boldsymbol{\alpha}$ values act on all of the layers of the multiview network simultaneously. In Figure \ref{Fig:fao_sender_large}, we show these estimated sender and receiver effects, for the $5$ largest countries (by population) in the EU. 
The relation of these latent positions with respect to the center of the latent space (i.e. the point $(0,0)$) is critical, since the predictors variables are determined by $\boldsymbol{\alpha}_{ij}'\textbf{z}$. An ellipsis which is concentrated close to the center of the space will tend to have low values of predictors, hence fewer connections. By contrast, an ellipsis reaching further areas will tend to send or receive more connections. The shape of the ellipsis may also be quite meaningful: a narrow shape will indicate that the corresponding country can only align with the latent positions of a few traded products, hence this might indicate that the country primarily specializes in importing or exporting some specific products.

Looking more closely at the plots in Figure \ref{Fig:fao_sender_large}, we see that not too many differences arise between countries with regard to the way they export products. The relatively large dispersion of the $\boldsymbol{\alpha}_{i\cdot}$ values signals that countries exhibit different behaviors on different products. The receiver plot instead highlights some stark differences between the countries, whereby the dispersion is lower but positioning is more diverse. This justifies the very different trading behaviors as measured by the number of received connections (i.e. the number of different products imported). For example, the fact that Germany is positioned so far from the center of the latent space signals that this country has a strong tendency to import most products, and especially those products that tend to align with the vertical axis.

\begin{figure}[hbtp]
\begin{center}
\includegraphics[width=0.4\textwidth, height=0.275\textheight]{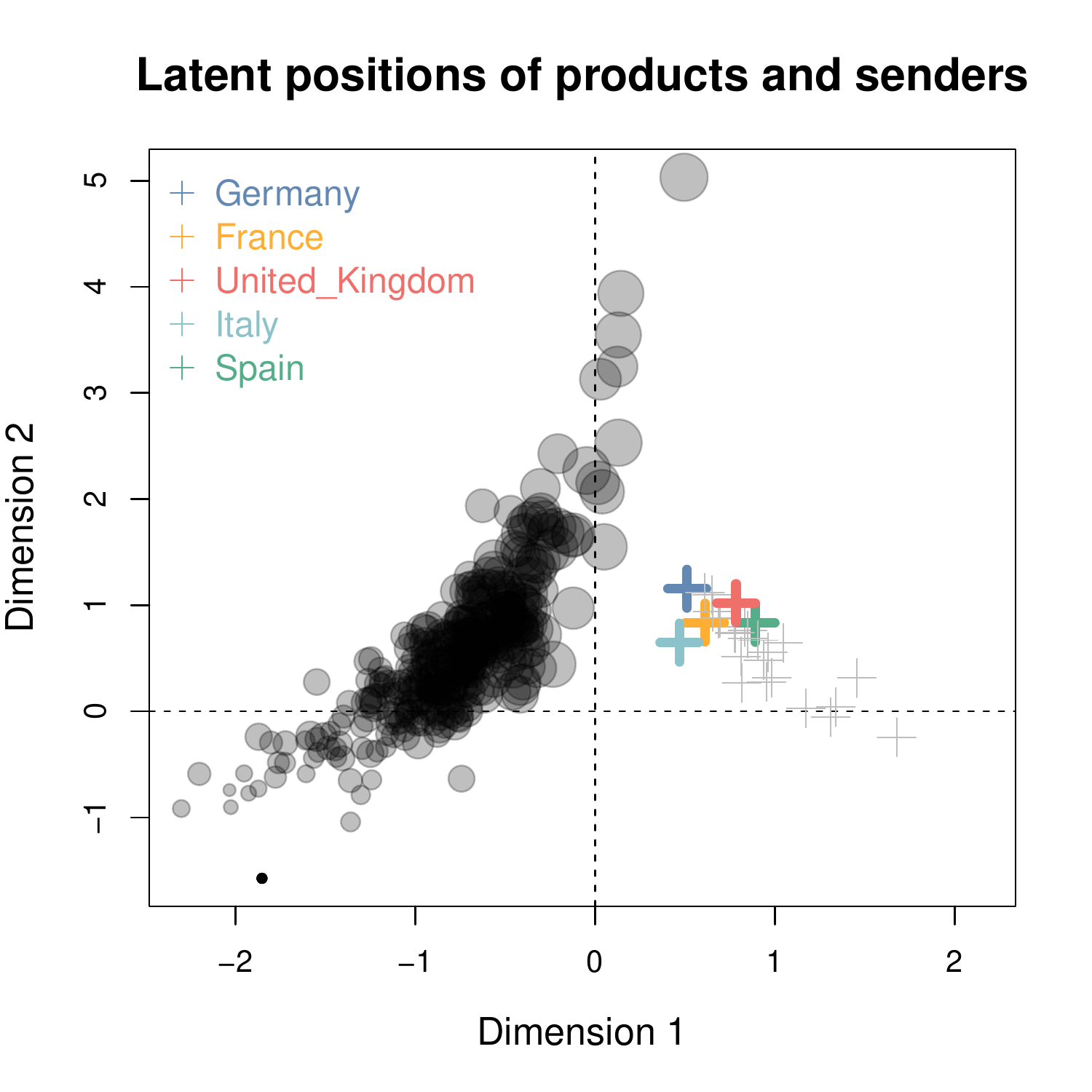}
\includegraphics[width=0.4\textwidth, height=0.275\textheight]{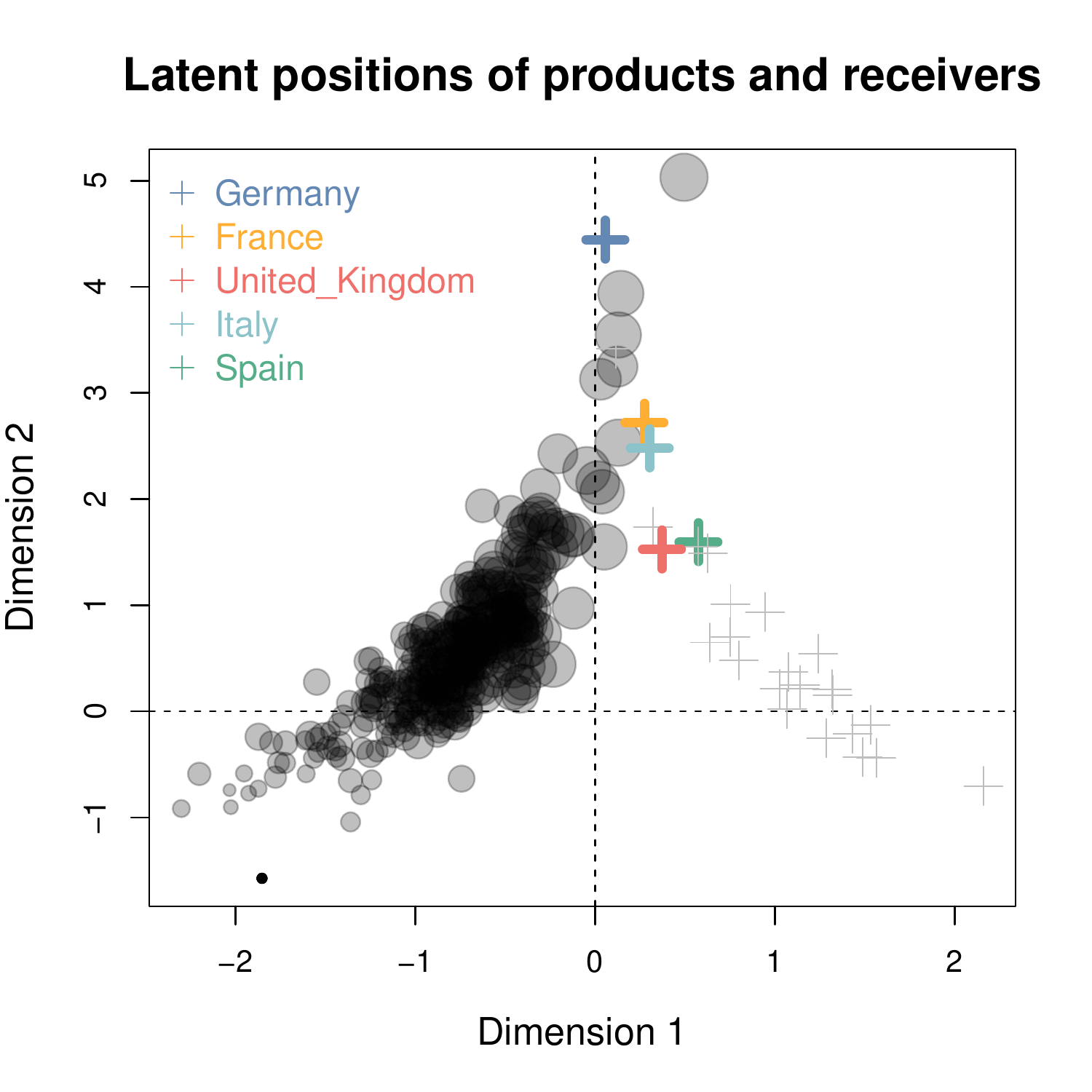}
\caption{FAO network. The circles are the latent positions of the estimated $\textbf{Z}$ (in black color), whereas the crosses are the median $\boldsymbol{\alpha}_{i\cdot}$ (left panel) and the median $\boldsymbol{\alpha}_{\cdot j}$ (right panel), for all $i$s and $j$s. The size of the circles indicate the importance of the product (total import + export). }
\label{Fig:fao_alpha_z}
\end{center}
\end{figure}

\begin{figure}[hbtp]
\begin{center}
\includegraphics[width=0.42\textwidth, height=0.275\textheight]{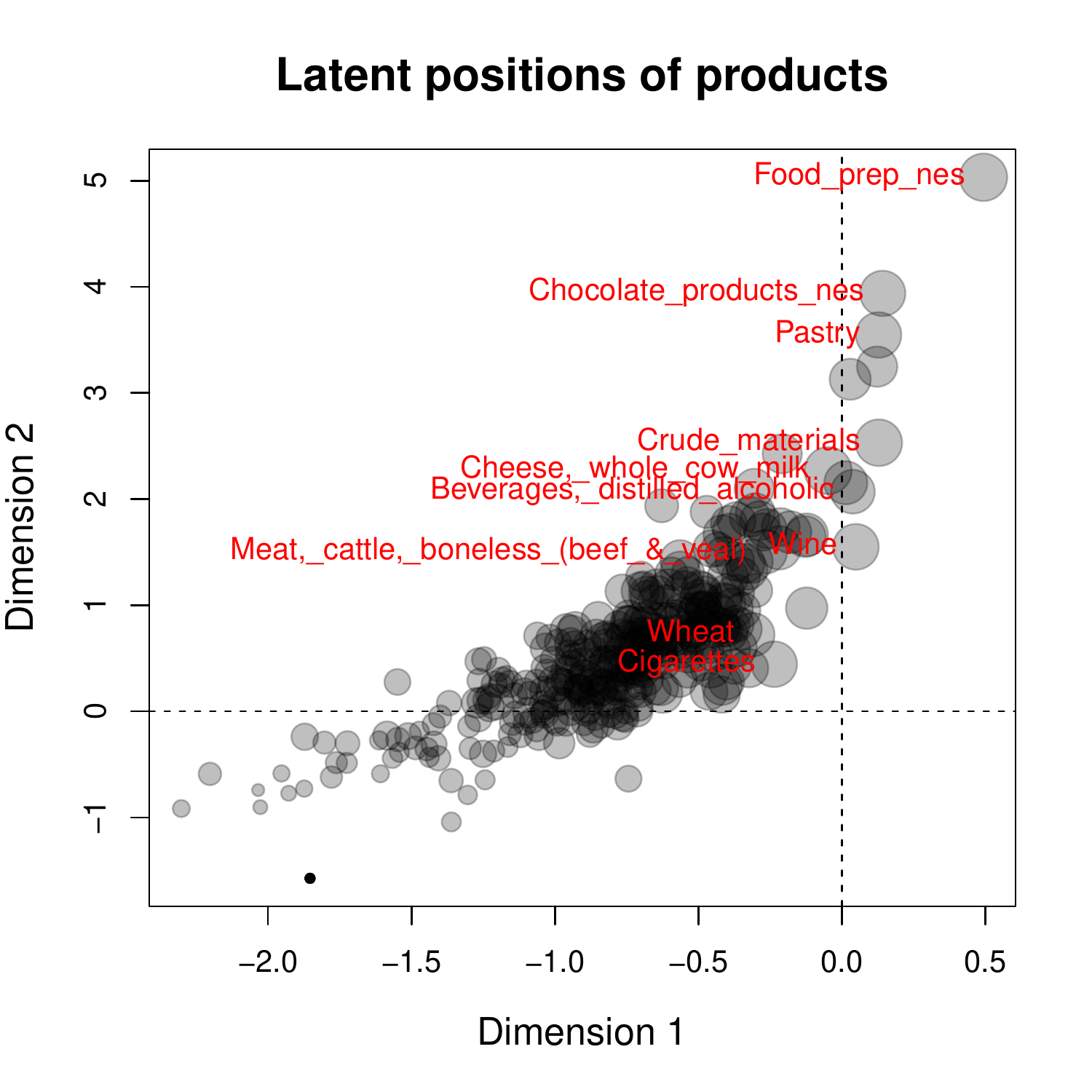}
\includegraphics[width=0.42\textwidth, height=0.275\textheight]{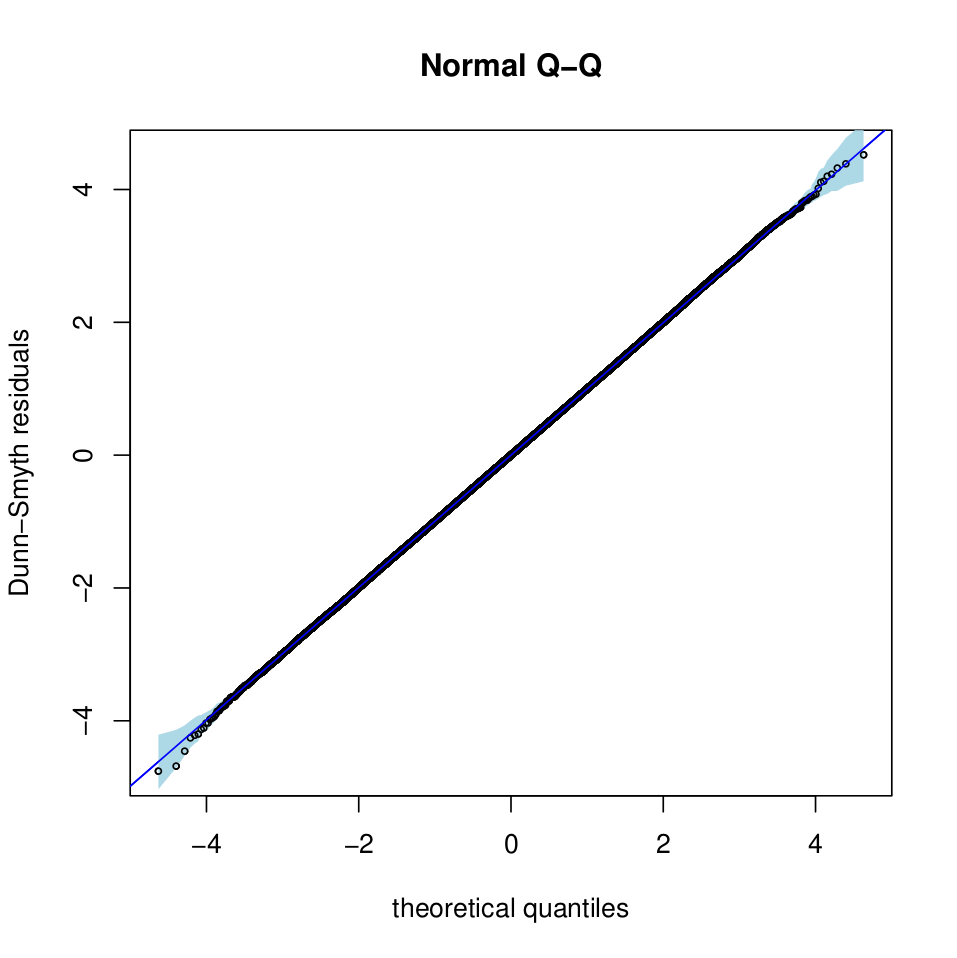}
\caption{FAO network. The left panel shows the estimated latent factor $\textbf{Z}$ for the products, with dot size representing their importance. The right panel shows the normal Q-Q plot for the Dunn-Smyth (randomized) residuals: shadowed area represents  a simulated point-wise 95\% confidence interval envelope.}
    \label{Fig:ggllvm_fao_z_vs_export}
\end{center}
\end{figure}

The analysis of the estimated latent factors  completes the picture, providing information on the products---indeed, we have different estimated values for each layer. 
We  propose a joint plot of estimated $\textbf{Z}$ along with the median sender and median receiver effects, in Figure \ref{Fig:fao_alpha_z}.
For the sender and receiver effects, we use the median value to avoid the effect of some outlier values.
This plot clearly shows that the receiver position of Germany strongly aligns with the top-traded products, since this country would import a large variety of them. Similar arguments can be made for the other large EU countries, although their position closer to the center would signal that they trade a smaller variety of commodities. Finally, at the lower ends of the distributions, we see that some EU countries (on the right side) are located literally opposite to the least traded products. This implies that these countries do not trade those products at all, as the corresponding predictors would be the lowest. Moreover, we show the estimated $\textbf{Z}$ for all products in the left panel of Figure \ref{Fig:ggllvm_fao_z_vs_export}.  
In the plot, we use as additional information the importance of each traded product, calculated as the total volume that is traded between all countries. The labels for the $10$ most traded products are shown on the plot. Each dot corresponds to a different product, and the size of each dot indicates its importance. We can see that the products are positioned roughly along a curved line, where the two ends correspond to the least and most important products. This is reasonable since the importance of a product is directly related to how many pairs of countries would trade it: this is the main feature that this latent space captures. There is an appreciable variability along this line, to indicate that products of same importance can be traded in different ways.

To conclude our analysis,  we consider the Dunn-Smyth (DS) residuals (also called  randomized quantile residuals, see \citet{DS98}), which are routinely-applied in the GLLVM  (and in GLM, as well) literature as a diagnostic tool: the distribution of the DS residuals converges to a standard normal if the model parameters are consistently estimated. To check this aspect, in Figure  \ref{Fig:ggllvm_fao_z_vs_export} (right panel) we display the normal Q-Q plot for the DS residuals as obtained using our GLAMLE. The picture  illustrates clearly that the quantiles of the residuals are close to the theoretical quantiles of a Gaussian random variable, with most of the dots laying on the 45-degrees line---even considering the point-wise 95\% envelope, the DS residuals look normally distributed, with no outlying values in the tails. This  diagnostic tool confirms that our model provides a good fit for the multiview network data.

\section{Conclusion and outlook} \label{Sec: concl}
We propose a novel inference technique for modeling dyadic variables in network data. To conduct inference, we derive a Laplace approximation to the  likelihood function for a GGLLVM for dyadic dependence. We study the properties of the proposed estimator and we exemplify its use via synthetic and real data. For the sake of illustration, we focus our applications on the binary case (Bernoulli random variables). Other applications to discrete distributions belonging to the exponential family (e.g. Poisson for count data), continuous distributions (e.g. Gaussian for real valued data) can be easily accommodated by our theory and method. As a direct extension of our method, we conjecture that it is able to deal also with mixture of discrete and continuous distributions (e.g. some $Y_{ij}$  follow a Bernoulli, some others a Poisson and the remaining follow a Gaussian distribution). The theory and method for GLAMLE in the presence of mixed variables should be developed adapting to our graph setting the results available in \citet{BKM11}, Ch. 7. This aspect is an interesting inference feature that, to our knowledge, other LVMs for networks do not have. 
Beside this direct extension,  this paper has the potential to trigger other novel research directions. We mention some of them.

(a) An interesting aspect  concerns with 
the selection of the number of latent dimensions.  Some approaches that have been considered for network models  include a Bayesian Information Criterion \citep{HRT07}, a Deviance Information Criterion \citep{Friel20166629}, shrinkage priors \citep{Durante20171547, rastelli2018sparse}.
In our framework, we may think of adapting the goodness-of-fit (GFI) test defined in \citet{Conne_2010}. The central idea is to propose a GFI which is based on the  comparison of some distance among the estimated latent variables and the corresponding distance among the manifest variables. 
We are planning to adapt this idea to our GGLLVM setting, as we did for the estimation via the GLAMLE. However, at the current stage of our research, it is not clear how to implement the GFI (bootstrap and bagging procedures are required) and additional investigations are needed.

(b) The derivation of the asymptotic properties of our estimator can yield further developments. In particular,  we plan to study the behavior of the Laplace approximated likelihood when the number of nodes grows to infinity at a certain rate, which is a function of the growing number of network views. We conjecture that this research direction will require to approach the asymptotic problem using graphons theory and the limits of dense graph sequences; see \citet{Lo12}. 

(c) In many real data analyses an interpretation of the latent variables is needed.  To this end, we are planning to explore the possibility of finding observable variables which are highly correlated to the filtered latent variables---indeed, the proposed Laplace approximation yields automatically estimates $\hat{\mathbf{z}}^{(k)}$, for each $k$-th network view. This should allow the use of some observables as proxies for the $\mathbf{Z}$, with the aim being to gain additional understanding on the underlying probability mechanism that drives the edges behavior. 

\color{black}
\small{
\bibliographystyle{apalike}
\bibliography{biblio}}

\newpage

\begin{footnotesize}
\appendix

\input{AppendixA}

\input{AppendixB}

\input{AppendixMCMC}

\end{footnotesize}

\end{document}

%% file: AppendixA.tex
\section{Constraining the GLAMLE for identifiability purposes} \label{Sec: uniq}

The estimating equations which define the GLAMLE  have multiple solutions. To see this, let us recall that our GGLLVM is based on a generalized linear model, so $\eta\{ E[Y_{ij}\vert \mathbf{Z}] \} = 
\boldsymbol{\alpha}'_{ij}\mathbf{Z} = \alpha_{ij,0} + \boldsymbol{\alpha}_{ij(2)} \mathbf{Z}_{(2)}$, where $\eta$ is the canonical link function and $\mathbf{Z}
_{(2)}$ are centred and standardized. Let
$\boldsymbol{O}$ be an orthogonal matrix of dimension $q \times q$. It is possible to rotate the matrix $\boldsymbol{\alpha}_{(2)}=\{\boldsymbol{\alpha}_{ij(2)} \}$ 
premultiplying 
it by $\boldsymbol{O}$ and thus obtaining a new matrix of parameters $\boldsymbol{\tilde\alpha}_{(2)} = \boldsymbol{O} \boldsymbol{\alpha}_{(2)}$. Since $ \mathbf{Z}
_{(2)}$ 
is centred and standardized and $\boldsymbol{O}$  is orthogonal, we define $\mathbf{\tilde{Z}}_{(2)} = \boldsymbol{O} \mathbf{Z}_{(2)}$ which is a new vector having 
a multivariate standard normal density. Moreover, the rotation induced by $\boldsymbol{O}$ does not change the product 
$\boldsymbol{\alpha}'_{(2)}\mathbf{Z}_{(2)}$, since
$ (\boldsymbol{\alpha}^{*}_{(2)})' \mathbf{Z}^{*}_{(2)} = \boldsymbol{\alpha}'_{(2)} \boldsymbol{O}' \boldsymbol{O} \mathbf{Z}_{(2)} = \boldsymbol{\alpha}'_{(2)}\mathbf{Z}
_{(2)}$.
Therefore, a rotation of $\boldsymbol{\alpha}$ gives a new matrix of parameters which is also a solution for the same model: the original and the rotated solutions are observationally equivalent. 

To fix the issue in \texttt{R}, one may decide to select a specific rotation using e.g. a \texttt{varimax} or a \texttt{promax} procedure on the estimated factor loadings. This is the solution routinely adopted in factor analysis. This solution has several advantages: it yields a simplification of the estimated quantities; it facilitates loadings and factors interpretation; it is computationally convenient since it is already implemented in the the statistical software. As an alternative and faster option, one may decide to impose some constraints on the orthogonal matrix. Specifically, we recall that
an orthogonal $q \times q$ matrix has $q (q - 1)/2$ degrees-of-freedom: the
matrix needs at least $q (q - 1)/2$ constraints on its elements to be unique. This represents the number of constraints that we must impose to obtain a unique solution to (\ref{Eq. GLAMLEshort}). Thus, as in Proposition 1 of \citet{HRVF04}, we have that if all the elements of the upper triangle of $\boldsymbol{\hat\alpha}_{(2)}$ are 
constrained, then $\boldsymbol{\hat\alpha}_{(2)}$ is completely determined, except for the sign of each column. If at least one constraint of the $\iota$-th column, with 
$\iota = 2,..., q$, is different from zero, then the sign of the corresponding column is determined. In the numerical studies of this paper, for both synthetic and real-data, we implemented this type of solution, exploiting the routine already available in \texttt{gllvm.TBM}.

%% file: AppendixB.tex
\section{Count data (Poisson random variables)} \label{Section: AppB}

\subsection{Example of count data}

In addition to binary data, also {count data} are often encountered in network analysis. For instance, the Internet service providers (ISPs) have a vested interest in monitoring the traffic on the Internet networks they maintain; see 
\citet{K09}. One fundamental quantity that ISPs look at is the volume of data sent from each origin node in the network to each destination node. A particularly interesting quantity is  the number of bytes flowing between an origin and a destination in, say, a given time frame (e.g. a five-minute interval). If one organizes such traffic flows in a matrix form, say with origins for rows and destinations for columns, the result is what the Internet measurement community calls a traffic matrix $\mathbf{Y}$. It is common to represent a given network by a graph, and assume that all vertices can serve as both origins and destinations. Then, 
each entry $Y_{ij}$ of the matrix represents count of bytes (e.g. $i \to j$). Repeated observations of $\mathbf{Y}$
(for instance for different communities) define the multiple network view setting.
Although in the Section \ref{Sec: application} we illustrate in detail the results for the binary version of the FAO data, here we give the key equations for count data as well.\\

\subsection{GLAMLE: key equations}
Consider the case where $\{Y_{ij}\}$ represents counts data for the edge $(ij)$. 
In this case, the $Y_{ij}$s can be modeled as
Poisson random variables. We write the Laplace-approximated likelihood under \textbf{A2}; the other case can be dealt with similar computation as in the binary data framework.

For a random sample $\boldsymbol Y^{(1)}, \boldsymbol Y^{(2)}, \dots, \boldsymbol Y^{(K)}$, the Laplace-approximated likelihood is:
\beq \label{Eq: LA_Poi}
\tilde{\ell}(\boldsymbol{\alpha})=\sum_{k=1}^{K} \l(-\frac{1}{2}\log\l[\det\l\{\Gamma\l(\boldsymbol{\alpha},\hat{\mathbf{z}}^{(k)}  \r)\r\}\r] +  
\sum_{i\ne j}^{n_V}\l\{Y_{ij}^{(k)}\boldsymbol{\alpha}'_{ij} \hat{\mathbf{z}}^{(k)}- \exp\l(\boldsymbol{\alpha}'_{ij}\hat{\mathbf{z}}^{(k)}\r) - \log Y_{ij}^{(k)} \r\}-
\frac{\hat{\mathbf{z}}^{(k)'}_{(2)}\hat{\mathbf{z}}^{(k)}_{(2)}}{2}\r )
\eeq
where $\hat{\mathbf{z}}^{(k)}=(1, (\hat{\mathbf{z}}_{(2)}^{(k)})')'$ is the 
root of $\partial Q\l(\boldsymbol{\alpha},\mathbf{z},\boldsymbol{Y}^{(k)}\r)/\partial \mathbf{z}= \mathbf 0$, with function
\beq \label{Eq: Q_Poi}
Q\l(\boldsymbol{\alpha},\mathbf{z},\boldsymbol y^{(k)}\r)= m^{-1}\l[  \sum_{i\ne j}^{n_V}\l\{y_{ij}^{(k)}\boldsymbol{\alpha}'_{ij} \mathbf{z}- 
\exp\l(\boldsymbol{\alpha}'_{ij}\mathbf{z}\r) - \ln y_{ij}^{(k)} \r\}-\frac{\mathbf{z}'_{(2)}\mathbf{z}_{(2)}}{2}-\frac{q}{2}\log(2\pi)\r].
\eeq
Estimation of the model parameter is obtained by solving the estimating equations yielded by the first order conditions, as in the case of binary variables.

%% file: AppendixMCMC.tex
\section{MCMC implementation details} \label{app:mcmc}

\subsection{Metropolis-within-Gibbs sampler}

Let $\mathcal{Y}$ represent the binary dyads of a multiview network, with $n_V$ nodes and $K$ layers. Let $\boldsymbol{\alpha}$ be a collection of $n_V \times n_V \times Q$ model parameters, and $\mathcal{Z}$ be a collection of $K \times Q$ model parameters. Here, $Q$ denotes the number of dimensions for the latent space.

The posterior distribution of our GGLLVM can be written as:
\begin{equation}
 p\left( \boldsymbol{\alpha}, \mathcal{Z} \middle \vert \mathcal{Y} \right) \propto p\left( \mathcal{Y} \middle \vert \boldsymbol{\alpha}, \mathcal{Z} \right) p\left( \boldsymbol{\alpha} \right) p\left( \mathcal{Z} \right)
\end{equation}
thanks to Bayes' theorem. Our goal is to obtain an empirical sample of $\boldsymbol{\alpha}$ and $\mathcal{Z}$ from their posterior distribution using a Metropolis-within-Gibbs sampler. Differently from the Laplace approach, we do not need to integrate out any parameters at this stage, since we can just discard the marginal samples, if needed.

As priors for the model parameters, we choose independent standard Gaussians. 
This means that 
$$
\alpha_{ijq} \stackrel{IID}{\sim} \mathcal{N}(0,1) \quad \quad \quad z_{kq} \stackrel{IID}{\sim} \mathcal{N}(0,1) \quad \quad \quad \forall i,j,k,q
$$

As regards the model's complete likelihood, this can be written as:
$$
\mathcal{L}_{\mathcal{Y}}\left( \boldsymbol{\alpha}, \mathcal{Z} \right) = \prod_{i,j,k} \sigma\left( \boldsymbol{\alpha}_{ij}'\textbf{z}_k \right)^{y_{ijk}} \left[ 1 - \sigma\left( \boldsymbol{\alpha}_{ij}'\textbf{z}_k \right) \right]^{1-{y_{ijk}}}
$$
where $\sigma$ is the sigmoid function $\sigma(x) = \frac{1}{1+e^{-x}}$, for any real valued $x$.

The sampler loops through each of the parameters in turn\footnote{Note that, diffently from other samplers used for latent position models, we do not use joint updates for any of the model parameters.} by sampling a value from their full-conditional distributions, which read as follows:
\begin{equation}
\begin{split}
 & p\left( \alpha_{ijq} \middle \vert \dots \right) \propto p\left( \alpha_{ijq} \right) \prod_{k} \sigma\left( \boldsymbol{\alpha}_{ij}'\textbf{z}_k \right)^{y_{ijk}} \left[ 1 - \sigma\left( \boldsymbol{\alpha}_{ij}'\textbf{z}_k \right) \right]^{1-{y_{ijk}}} \\
 & p\left( z_{kq} \middle \vert \dots \right) \propto p\left( z_{kq} \right) \prod_{i,j} \sigma\left( \boldsymbol{\alpha}_{ij}'\textbf{z}_k \right)^{y_{ijk}} \left[ 1 - \sigma\left( \boldsymbol{\alpha}_{ij}'\textbf{z}_k \right) \right]^{1-{y_{ijk}}} \\
\end{split}
\end{equation}
where the dots indicate the list of all other model parameters and the data.

Since the full-conditional distributions are not in standard form, random walk Metropolis steps are used. Given a model parameter $\theta$, we draw a new candidate $\tilde\theta = \theta + \varepsilon$, where $\varepsilon$ is drawn from a standard Gaussian proposal distribution. Then, we calculate the acceptance probability
$$
A = \min\left\{ 1\ ,\ \frac{p\left( \tilde{\theta} \middle \vert \dots \right)}{p\left( \theta \middle \vert \dots \right)} \right\}
$$
which we use to determine whether we update $\theta$ with $\tilde\theta$. 
Either way, the value of the model parameter will be appended to the corresponding posterior sample.

\subsection{Posterior sample interpretation}

Similarly to other latent position models \citep{HRH02, shortreed2006positional}, the MCMC procedure leads to a non-interpretable posterior sample. 
This is due to the fact that the likelihood function is invariant with respect to rotations and reflections of the latent space.
Since these transformation may inadvertly occur at any point during the sampling, we cannot summarize our posterior sample using typical estimators, such as the posterior mean.
However, we note that it may be possible to extract a meaningful maximum-a-posteriori estimator.

Due to the nature of our simulation study, this identifiability problem does not affect our results.
In fact, we only extract and calculate the connection probabilities $\boldsymbol{\pi}$ from our posterior samples.
These probabilities are themselves invariant to rotations and reflections of the latent space, hence by passing the issue.

Since our simulation study does not directly require point estimates of $\boldsymbol{\alpha}$ and $\mathcal{Z}$, we do not address this issue here. 
However, we note that Procrustes matching and other relevant approaches \citep{shortreed2006positional} may be adapted to our setup.